\documentclass[prd,twocolumn,tightenlines,showpacs,nofootinbib,amsfonts,amssymb,amsmath]{revtex4-1}
\usepackage{graphicx,color}
\begin{document}
\newcommand{\etal}{{\it et al.}}
\newcommand{\bx}{{\bf x}}
\newcommand{\bn}{{\bf n}}
\newcommand{\bk}{{\bf k}}
\newcommand{\dd}{{\rm d}}
\newcommand{\dslash}{D\!\!\!\!/}
\def\ga{\mathrel{\raise.3ex\hbox{$>$\kern-.75em\lower1ex\hbox{$\sim$}}}}
\def\la{\mathrel{\raise.3ex\hbox{$<$\kern-.75em\lower1ex\hbox{$\sim$}}}}
\def\beq{\begin{equation}}
\def\eeq{\end{equation}}

\leftline{UMN--TH--3429/15}

\vskip-2cm
\title{Trajectories with suppressed tensor-to-scalar ratio in Aligned Natural Inflation}
\author{Marco Peloso and Caner Unal
}
\affiliation{
 School of Physics and Astronomy,
University of Minnesota, Minneapolis, 55455 (USA)\\
}
\vspace*{2cm}

\begin{abstract}

In  Aligned Natural Inflation, an alignment between different potential terms produces an inflaton excursion greater than the axion scales in the potential.  We show that, starting from a general potential of two axions with two aligned potential terms, the effective theory for the resulting light direction is characterized by four parameters: an effective potential scale, an effective axion constant, and two extra parameters (related to ratios of the axion scales and the potential scales in the $2-$field theory). For all choices of these extra parameters, the model can support inflation along valleys (in the $2-$field space) that end in 
 minima of the potential. This leads to a phenomenology similar to that of single field Natural Inflation. For a significant range of the extra two parameters, the model possesses also  higher altitude inflationary trajectories passing through  saddle points of the $2-$field potential, and disconnected from any minimum.  These plateaus end when the heavier direction becomes unstable, and therefore all of inflation takes place close to the saddle point, where - due to the higher altitude - the potential is flatter (smaller $\epsilon$ parameter). As a consequence, a tensor-to-scalar ratio  $r = {\rm O } \left( 10^{-4} - 10^{-2} \right)$ can be easily achieved in the allowed $n_s$ region, well within the  latest $1 \sigma$ CMB contours.

\end{abstract}
 \date{April 2015}
 \maketitle

\section{Introduction}
\label{sec:intro}

The Cosmic Microwave Background (CMB) data are in excellent agreement with the predictions of slow-roll inflation \cite{Ade:2015lrj}. Models that lead to sufficiently long inflation are very sensitive to Planck-scale physics. Even higher dimension Planck-Mass suppressed operators, such as $\phi^2 \, V / M_p^2$,~\footnote{In this expression, $\phi$ denotes the inflaton, $V$ its potential, and  $M_p$ is the (reduced) Planck scale $M_p = \left( 8 \pi G_N \right)^{-1/2}$, where $G_N$ is Newton's contant.} induce corrections of ${\rm O } \left( 1 \right)$ or larger to the slow roll parameter $\eta$, and therefore shorten the duration of inflation to a few e-folds. This UV-sensitivity is present for any inflationary potential, but it particularly affects models that generate a relatively large gravity wave signal (tensor-to-scalar ratio $r \ga 0.01$), as in such models the  inflaton vacuum expectation value (vev) changes by an amount  greater than the Planck scale during inflation \cite{Lyth:1996im}.

A well known and simple symmetry that can forbid such operators is a shift symmetry, namely the invariance of the action under the transformation $\phi \rightarrow \phi + {\rm constant}$.  A field possessing this symmetry, at least at an approximate level, is denoted as an axion. In order to have a nontrivial $\phi-$dependent potential, the symmetry cannot be exact, but it can be broken at a technically naturally small level: assume that only one sector of the theory breaks the shift symmetry, leading to a nontrivial potential for $\phi$; if this potential is small, quantum corrections to it will also be small, as they must be proportional to it  (loops that do not contain such term do not break the shift symmetry, and therefore do not generate a potential for the axion). The first work that recognized the merits of employing a shift symmetry in inflation is   \cite{Freese:1990rb}, where the resulting class of models was named {\it Natural Inflation}. 

A relevant scale in axion inflation is the  axion scale $f$.  This is the scale that determines the least irrelevant shift-symmetric coupling, often the dimension five coupling $\frac{\phi}{f} F {\tilde F}$ to gauge fields. In presence of this operator, gauge instantons  break nonperturbatively the continuous shift symmetry down to the discrete  $\phi \rightarrow \phi + 2 \pi f$. Therefore, $f$ controls the periodicity of the axion potential, in absence of any explicit breaking of the shift-symmetry. Refs.  \cite{Freese:1990rb,Adams:1992bn} studied the simplest potential with such property
\begin{equation}
V = \Lambda^4 \left[ 1 - \cos \left( \frac{\phi}{f} \right) \right] \;. 
\label{natural}
\end{equation} 
As conventionally done in the literature, in this work we use the term Natural Inflation exclusively for (\ref{natural}), but we 
credit  \cite{Freese:1990rb,Adams:1992bn} for realizing the relevance of the shift-symmetry protection, and not simply of the potential (\ref{natural}). 

Models of axion inflation are under control for $m_\phi, \, H < f$, where $m_\phi$ denotes the axion mass, and $H$ the Hubble rate during inflation. This is due to the fact that the theory of the axion is obtained by integrating out modes that are heavier than $f$. Ideally, one would also want to have $f < M_P$ (see  \cite{Pajer:2013fsa} for a partial list of relevant literature on this issue). One reason for this is that one expects quantum gravity to break the shift symmetry, as any global symmetry, at the Planck scale. This argument does not apply if the shift symmetry originates from a local symmetry, as  typically in string theory. However, all known controlled strong theory constructions have sub-Planckian axion scales. The 
 $f < M_p$ request contrasts with the fact that the model  (\ref{natural}) is consistent with the CMB results only for  $f \ga 7 \, M_p  $\cite{Ade:2015lrj}. Due to the more speculative nature of these arguments, the conservative approach of relaxing the $f < M_p$ requirement can be justifiable. Nonetheless, a large number of studies have been appeared in the recent literature, with the goal of providing a UV completion of natural inflation, formulated in terms of axions with a sub-Planckian scale. 

Mechanisms developed in these works include: identifying the axion with the extra component of a gauge field in a $5$d theory compactified on a circle \cite{ArkaniHamed:2003wu,ArkaniHamed:2003mz,Kaplan:2003aj,Paccetti:2005zm}; using two axions with a nearly aligned potential  \cite{Kim:2004rp}, using $N = {\rm O } \left( 10^2 - 10^3 \right)$ axions \cite{Dimopoulos:2005ac} in the mechanism of assisted inflation  \cite{Liddle:1998jc}; generating a monodromy through an explicit breaking of the shift symmetry  \cite{Silverstein:2008sg,McAllister:2008hb,Hannestad:2009yx,Marchesano:2014mla,Harigaya:2014eta,McAllister:2014mpa,Li:2014vpa,Blumenhagen:2014nba,Minor:2014xla,Hebecker:2014kva,Flauger:2014ana,Li:2014unh,Garcia-Etxebarria:2014wla,Blumenhagen:2015qda}; slowing down the inflaton from particle production,  \cite{Mohanty:2008ab,Anber:2009ua,Visinelli:2011jy,Albrecht:2014sea} (as in the original model of warm inflation \cite{Berera:1995ie}); coupling the axions to higher forms  \cite{Kaloper:2008fb,Kaloper:2011jz,Dudas:2014pva}; considering a non-standard mixing of the axion with gravity \cite{Germani:2010hd}; modifying the axion evolution with a coupling to a non-abelian vector field with nonvanishing vev  \cite{Adshead:2012kp},~\footnote{A related mode, that can be understood as the limit of \cite{Adshead:2012kp} in a certain parameter range, has been formulated in  \cite{Maleknejad:2011jw}. Both the original models of  \cite{Adshead:2012kp} and \cite{Maleknejad:2011jw} are ruled out by either stability considerations or the CMB data  \cite{Dimastrogiovanni:2012ew,Adshead:2013qp,Namba:2013kia}. See also ref. \cite{Obata:2014loa} for a recent extension of \cite{Adshead:2012kp}  to two axions.} considering higher derivative terms \cite{Ohashi:2012wf,Maity:2012dx}; considering multiple sinusoidal functions \cite{Czerny:2014wza,Czerny:2014xja,Kallosh:2014vja}; realizing an alignment from kinetic mixing \cite{Bachlechner:2014hsa,Shiu:2015uva,Shiu:2015xda};  explicitly breaking the shift-symmetry via $\phi^m F {\tilde F}$ operators \cite{McDonald:2014nqa,McDonald:2014rha}; employing  Coleman-Weinberg contributions to the axion potential \cite{Croon:2014dma,Croon:2015fza}. Alternatively, models with non-compact axions have also recently been  considered \cite{Burgess:2014tja,Csaki:2014bua}. 

Very recently, a discussion has started in the literature of whether also models with sub-Planckian axions that produce an effective super-Planckian excursion (as the model of  \cite{Kim:2004rp} that we study here) are also affected by large quantum gravity corrections \cite{delaFuente:2014aca,Rudelius:2015xta,Montero:2015ofa,Brown:2015iha,Bachlechner:2015qja,Hebecker:2015rya}.  An argument in this discussion is that fulfilling the Weak Gravity Conjecture  (WGC)\cite{ArkaniHamed:2006dz} in such models implies the existence of gravitational instantons that produce unsuppressed correction to the potential. The conditions under which such corrections are relevant, and their magnitude, are still under debate. In particular, ref.  \cite{Bachlechner:2015qja} discussed conditions under which there can be gravitational instantons  that fulfill the WGC, without introducing unsuppressed higher harmonics to the potential. We do not address these aspects in the present work, but we rather  study the phenomenology of the aligned model of  \cite{Kim:2004rp} taken at face value. Our goal is to show that the model possesses inflationary trajectories that have not been emerged in the many studies of the model, and that have a particularly interesting phenomenology (as, contrary to the predictions obtained from (\ref{natural}), they are compatible with the CMB data at $1 \sigma$). Although our computations are restricted to  \cite{Kim:2004rp}, we hope that inflationary solutions with analogous properties can also be present in other models of multi-field axion inflation, and that our result can motivate the search for such solutions. 

Prior to the present work, the alignment mechanism of  \cite{Kim:2004rp} was  studied in \cite{Tye:2014tja,Kappl:2014lra}, and extended to the $N>2$ axions case in  \cite{Chatzistavrakidis:2012bb,Choi:2014rja,Higaki:2014pja,Higaki:2014mwa,Choi:2014xva}. The works \cite{Berg:2009tg,Ben-Dayan:2014zsa,Long:2014dta,Gao:2014uha,Li:2014lpa,Ben-Dayan:2014lca,Carone:2014cta,Ali:2014mra,Bachlechner:2014gfa,Kappl:2015pxa,Ruehle:2015afa}  studied the embedding of the alignment mechanism in string theory.~\footnote{Other studies of axion inflation in string theory can be found in \cite{Kallosh:2007ig,McAllister:2007bg,Baumann:2014nda,Blumenhagen:2014gta,Hebecker:2014eua,Grimm:2014vva,Kenton:2014gma,Westphal:2014ana,Abe:2014pwa,Mazumdar:2014qea,Chernoff:2014cba}. For a generic study of inflationary models based on alignment, see \cite{Burgess:2014oma}.} Despite this extensve series of works,  a detailed study of the phenomenological predictions of  Aligned Natural Inflation  \cite{Kim:2004rp} is still missing. This is the goal of the present work. 

The most general potential for the two axions introduced in \cite{Kim:2004rp} is characterized by two potential terms and four axion scales. An alignment between these scales results in  a hierarchy between the mass scales of the two axions. We denote the light and heavy state by $\phi$ and $\psi$, respectively. Despite the simplicity of the model, 
obtaining the phenomenology  for all possible choices of the parameters requires extensive computations. We improve over existing discussions of the mechanism  by computing the inflationary trajectories  $\psi \left( \phi \right)$ in the model, and the effective $1-$field potential along such trajectories. We find that, in the limit of strong alignment, this effective potential is characterized by four parameters: an effective axion scale $f_\phi$, an overall potential scale $\Lambda^4$, and two additional parameters, $r_f$ and $r_\Lambda$, related to ratios of the axion scales and the potential scales of the starting potential. For any fixed $r_f$ and $r_\Lambda$, the model has the same freedom as Natural Inflation: the value of $f_\phi$ controls the overall flatness of the potential (and so the amount of inflation along an inflationary trajectory), while the scale $\Lambda$ is fixed by the normalization of the scalar perturbations. Therefore,  we learn that, to study the general phenomenology of the Aligned Natural Infaltion, we ``only'' need to vary over the combinations  $r_f$ and $r_\Lambda$, in addition to what is done for the single field model of Natural Inflation. 

An inflationary trajectory $\psi \left( \phi \right)$ needs to have (i) sufficient flatness along the light direction, and (ii) stability along the heavy direction, $\frac{\partial^2 V}{\partial \psi^2} > 0$. From our analysis we learn that the trajectories 
of Aligned Natural Inflation behave in  qualitatively different ways according to the relative value of $r_\Lambda$ vs $r_f$. For all choices of these two parameters, the model admits inflationary trajectories connected to minima of the potential: the heavy direction is stable along these trajectories, and inflation ends close to the minimum, when the $\phi$ direction is no longer flat. These trajectories lead to a phenomenology analogous to that of Natural Inflation. As for this model, the predicted values of $n_s$ and $r$ lie on a region which is outside the latest $1 \, \sigma $ CMB contours \cite{Ade:2015lrj}. For a significant range of the extra parameters (specifically, for $r_\Lambda \epsilon \left[ \frac{1}{r_f^4} ,\,  
 \frac{1}{r_f^2} \right]$ in the case of $r_f>1$, and for  $r_\Lambda \epsilon \left[ \frac{1}{r_f^2} ,\,  
 \frac{1}{r_f^4} \right]$ in the case of $r_f<1$) we  find the presence of a new class of trajectories that has not emerged in previous discussions of the mechanism. This new class of trajectories occurs near a saddle point of the full potential, and inflation terminates not because a minimum has been reached, but because the heavy direction becomes unstable.~\footnote{This behavior is reminiscent of hybrid inflation \cite{Linde:1993cn}.} Thanks to this feature, the flatness of these trajectories along the light direction is decoupled from the end of inflation, and inflation occurs on a higher altitude plateau, where the potential is flatter. This leads to a much smaller value of $r$ with respect to Natural Inflation,   well within the $1 \sigma$ CMB bounds. 

Another characteristic of models of axion inflation is that  the shift symmetry highly constrains the coupling of the inflaton to matter. One expects that the dominant among such interactions is the $\frac{c}{f} \phi F {\tilde F}$ coupling to gauge fields, where $c$ is a dimensionless coefficient that in the spirit of effective field theories is expected to be of ${\rm O } \left( 1 \right)$. As first shown in \cite{Barnaby:2010vf}, a coupling $\frac{c}{f} \ga \frac{1}{10^{16} \, {\rm GeV}}$ of the inflaton to any light gauge field results in overproduction of gauge quanta during inflation. In the limiting case  $\frac{c}{f} = {\rm O } \left( \frac{1}{10^{16} \, {\rm GeV}} \right)$ such a production can have interesting phenomenological consequences, that we review in Section  \ref{sec:phiFF}. While this production is negligible in Natural Inflation (since $f \ga 7 \, M_p$ is required), a large enough coupling can be easily present in Aligned Natural Inflation. We update the discussion of this effect already presented in the literature by computing the precise value of the coupling necessary to obtain a sizable effect along some of the inflationary trajectories obtained in this work. 

\bigskip

The paper is organized as follows. In Section  \ref{sec:nkp-review} we review the model introduced in \cite{Kim:2004rp}, and we obtain the effective $1-$field potential in the limit of strong alignment. In Section \ref{sec:trajectories} we discuss more in details the light field trajectories in the $2-$field potential. In Section  \ref{sec:pheno} we study the resulting phenomenology in the $\left\{ n_s - r \right\}$ plane.  In Section \ref{sec:phiFF} we discuss possible additional signatures of Aligned Natural Inflation that can emerge from the couplings of the axions to light gauge fields during inflation. In Section \ref{sec:conclusion} we present our conclusions. In Appendix \ref{sec:ana} we present some analytic properties of the inflationary trajectories. In Appendix \ref{sec:perts} we briefly review the computation of cosmological perturbations in  $2$-field inflationary models, that we use to obtain some of the results of Section \ref{sec:pheno}. In Appendix \ref{app:incorrect} we warn against an approximation that does not produce the correct $1-$field description of the model.

\section{Natural Aligned Inflation and the  strong alignment limit}
\label{sec:nkp-review}

The general $2-$field ($\theta ,\, \rho$) model of Natural Aligned Inflation is characterized by four axion constants and two potential scales \cite{Kim:2004rp}
\begin{equation}
V = \Lambda_1^4 \left[ 1 - \cos \left( \frac{\theta}{f_1} + \frac{\rho}{g_1} \right) \right] 
+  \Lambda_2^4 \left[ 1 - \cos \left( \frac{\theta}{f_2} + \frac{\rho}{g_2} \right) \right] \;. 
\label{knp}
\end{equation} 
This potential has one exact flat direction, which we denote by $\phi$, for  $f_1/g_1=f_2/g_2$. If this relation is only approximately valid, that direction becomes flatter than the naive expectation, namely $m_\phi \ll \frac{\Lambda_i^2}{f_i} \;,\;   \frac{\Lambda_i^2}{g_i}$. Equivalently, this can be seen as having an axion $\phi$ with an effective scale $f_\phi \gg f_i ,\, g_i$. This is the aligned mechanism of \cite{Kim:2004rp}.  

To study this model, we find covenient to express the two potential scales as 
\begin{equation}
\Lambda_1^4 \equiv \frac{1}{1+r_\Lambda} \, \Lambda^4 \;\;,\;\; 
\Lambda_2^4 \equiv \frac{r_\Lambda}{1+r_\Lambda} \, \Lambda^4 \;\;, 
\label{def-L-xi}
\end{equation}
so that $\Lambda$ is an overall potential scale, and $r_\Lambda$ is the ratio between the two scales  in (\ref{knp}). We also find convenient to express the ratio between the axion scales as
\begin{equation}
\frac{g_1}{f_1} \equiv \frac{r_g}{1+\alpha} \;\;,\;\; 
\frac{g_2}{f_2} \equiv \frac{r_g}{1-\alpha}  \;,  
\end{equation}
Finally, we redefine 
\begin{equation} 
f_1 \equiv r_f \, f \;\;,\;\; f_2 \equiv \frac{f}{r_f} \,. 
\label{def-f-rf}
\end{equation} 

In this way, we reformulate the model (\ref{knp}) in terms of two parameters of mass dimension one ($f$ and $\Lambda$) and four dimensionless parameters: 
\begin{eqnarray}
\frac{V}{\Lambda^4} &=& 1 - \frac{1}{1+r_\Lambda} \, \cos \left[ \frac{1}{r_f} \left( \frac{\theta}{f} + \frac{1+\alpha}{r_g} \, \frac{\rho}{f} \right) \right] \nonumber\\ 
&& \quad\quad- \frac{r_\Lambda}{1+r_\Lambda} \, \cos \left[ r_f \left( \frac{\theta}{f} +   \frac{1-\alpha}{ r_g} \, \frac{\rho}{f} \right) \right] \,.  
\label{V-start}
\end{eqnarray}
The potential is symmetric under $S_1 = \left\{ r_\Lambda \rightarrow \frac{1}{r_\Lambda} \;,\; \alpha \rightarrow - \alpha \;,\; r_f \rightarrow \frac{1}{r_f} \right\}$ (interchange of the two terms) and under $S_2 = \left\{ \theta \rightarrow - \theta  \;,\; \rho \rightarrow - \rho \right\}$ (the potential is even). We stress that (\ref{V-start}) is equivalent to (\ref{knp}), and that no approximation has yet been made.

In this work we obtain phenomenological results for Aligned Natural Inflation using the exact $2-$field model 
(\ref{V-start}) and the effective $1-$field model that we derive below. The excellent agreement between the results from the $2-$field and the $1-$field model confirms the accuracy of the approximation that we perform in the reminder of this Section. Specifically, we are interested in the model in the limit of strong alignment. In our parametrization, this means $\vert \alpha \vert \ll 1$. In the following, we study the theory to the  leading nontrivial order in $\alpha$.   We start by performing the rotation
\begin{eqnarray}
&& \left( \begin{array}{c} \theta \\ \rho \end{array} \right) = R \,  
\left( \begin{array}{c} \phi \\ \psi \end{array} \right) \,, \nonumber\\ 
&& R_{11} = R_{22} = \frac{1}{\sqrt{1+r_g^2}} + \alpha \, \frac{r_g^2}{\left( 1 + r_g^2 \right)^{3/2}} \frac{1-r_f^4 \, r_\Lambda}{1+r_f^4 \, r_\Lambda} \;, 
\nonumber\\ 
&& R_{12} = - R_{21} = \frac{r_g}{\sqrt{1+r_g^2}} - \alpha \, \frac{r_g}{\left( 1 + r_g^2 \right)^{3/2}}  \frac{1-r_f^4 \, r_\Lambda}{1+r_f^4 \, r_\Lambda} \,.  
\label{rotation}
\end{eqnarray} 
The rotation matrix is orthogonal  up to ${\rm O } \left( \alpha^2 \right)$ corrections, and therefore the kinetic term in the new basis reads 
\begin{eqnarray}
 \frac{ \left( \partial \theta \right)^2 }{2} + \frac{  \left( \partial \rho \right)^2  }{2} = 
\left[  \frac{ \left( \partial \phi \right)^2 }{2} + \frac{  \left( \partial \psi \right)^2  }{2} \right] \, 
\left[ 1 + {\rm O } \left( \alpha^2 \right) \right]  \;, 
\label{kinetic}
\end{eqnarray} 
and in the following we disregard the $ {\rm O } \left( \alpha^2 \right) $ corrections.  
 
The transformation  (\ref{rotation}) rotates (up to ${\rm O } \left( \alpha^2 \right)$ corrections) the fields $\theta$,\, $\rho$ in the light $\phi$ and heavy $\psi$ eigenstate of the potential (\ref{V-start}) expanded at quadratic order around the minimum located at the origin. Along the inflationary trajectories, $\psi$ is always the heavy field (up to ${\rm O } \left( \alpha \right)$ corrections), and the trajectories are obtained by integrating $\psi$ out. Integrating $\psi$ out leads to a trajectory $\psi = \psi_s \left( \phi \right)$, which has the property $\psi_s = {\rm O } \left( \phi^3 \right)$ next to the origin, but that is $\psi_s \neq 0$ in general. It is crucial to compute the precise trajectory $\psi_s \left( \phi \right)$ in order to obtain the correct $1-$field effective description of the model.~\footnote{In Appendix \ref{app:incorrect} we show explicitly through some examples that setting $\psi =0$ everywhere provides in general an incorrect CMB phenomenology.} 

Under the rotation (\ref{rotation}), the potential becomes 
\begin{eqnarray}
 \frac{V}{\Lambda^4} &=& 1 - \frac{ \cos \left( c_1 \, \alpha \, \phi + c_2 \, \psi \right) }{1+r_\Lambda} \,
 - \frac{r_\Lambda \;  \cos \left( c_3 \, \alpha \, \phi + c_4 \, \psi \right)  }{1+r_\Lambda} \,,  \nonumber\\ 
c_1 &\equiv& - \frac{2 \, r_f^3 \, r_\Lambda }{f \sqrt{1+r_g^2} \left( 1 + r_f^4 \, r_\Lambda \right)} 
 \, \left[ 1 + {\rm O } \left( \alpha \right) \right] \,, \nonumber\\ 
c_2 &\equiv& \frac{\sqrt{1+r_g^2}}{f \, r_f \, r_g}  \left[ 1 + {\rm O } \left( \alpha \right)  \right] \,, \nonumber\\ 
c_3 &\equiv&   \frac{2 \, r_f   }{f \sqrt{1+r_g^2} \left( 1 + r_f^4 \, r_\Lambda \right)} 
 \, \left[ 1 + {\rm O } \left( \alpha \right) \right] \,, \nonumber\\ 
c_4 &\equiv&  \frac{r_f \sqrt{1+r_g^2}}{ f \, r_g} \left[ 1 + {\rm O } \left( \alpha \right)  \right]  \,, 
\end{eqnarray} 
and in the following we disregard the ${\rm O } \left( \alpha \right)$ corrections to the coefficients $c_i$. This expression for the potential manifestly shows that $\phi$ is the light direction in the limit of large alignment, since it enters multiplied by $\alpha \ll 1$ and by an ${\rm O } \left( \frac{1}{f} \right)$  factor in the argument of the two cosines, while $\psi$ enters times an  ${\rm O } \left( \frac{1}{f} \right)$  factor without any $\alpha$ suppression.~\footnote{We are implicitly assuming that, apart from $\alpha$, the other dimensionless parameters in  (\ref{V-start}) are of order one.}  Therefore, apart from special points where $\frac{\partial^2 V}{\partial \psi^2} = 0$, the curvature of the potential in the $\psi$ direction is much greater than the curvature in the $\phi$ direction. 

To visualize and study the potential, it is convenient to perform a different rescaling of the $2-$fields: 
\begin{equation} 
{\tilde \phi} = \frac{2 \, \alpha  }{\sqrt{1+r_g^2}} \, \frac{ \phi }{ f } \equiv \frac{\phi}{f_\phi} \;\;\;,\;\;\; 
{\tilde \psi} = \frac{\sqrt{1+r_g^2}}{r_g} \, \frac{ \psi }{ f }  \equiv \frac{\psi}{f_\psi} \,, 
\label{tilde-fields} 
\end{equation} 
This allows us to draw a contour plot of the potential $V \left( {\tilde \phi} ,\, {\tilde \psi} \right)$  in which the curvature appears to be comparable in both directions, and the trajectories can be better visualized. However, one should keep in mind that, in the basis $\left\{ \phi ,\, \psi \right\}$ of the canonically normalized fields the curvature is much greater in the $\psi$ direction. As a consequence, starting from a generic initial condition, the system first evolves in the $\psi$ direction, with practically no change in $\phi$, until it reaches a  minimum $\frac{\partial V}{\partial \psi} = 0$. Then, the system continues to evolve along a trajectory that minimizes $V$ with respect to $\psi$. 

In terms of the rescaled fields (\ref{tilde-fields}),  the action of the model is 
\begin{eqnarray} 
&& {\cal L } =  \frac{f_\phi^2}{2} \, 
\left( \partial {\tilde \phi} \right)^2  + \frac{f_\psi^2}{2} \, \left( \partial {\tilde \psi} \right)^2  - \Lambda^4 \, V_2  \nonumber\\ 
&& V_2 \equiv 1- \frac{1}{1+r_\Lambda}   \cos \left( - \frac{r_f^3 \, r_\Lambda \, {\tilde \phi}}{1 + r_f^4 \, r_\Lambda} + \frac{\tilde \psi}{r_f} \right) \nonumber\\ 
&& \quad\quad\quad\quad \quad\quad  \quad 
- \frac{r_\Lambda}{1+r_\Lambda} \cos \left( \frac{r_f {\tilde \phi}}{1 + r_f^4 \, r_\Lambda} + r_f \, {\tilde \psi} \right) \;. 
\label{V2}
\end{eqnarray} 
This formulation of the theory is characterized by $5$ parameters, and it is equivalent to the original theory up to ${\rm O } \left( \alpha \right)$ corrections. The potential in (\ref{V2}) is still invariant under the  symmetries $S_{1,2}$ defined above, which now read  $S_1 = \left\{ r_\Lambda \rightarrow \frac{1}{r_\Lambda} \;,\; {\tilde \phi}  \rightarrow - {\tilde \phi} \;,\; r_f \rightarrow \frac{1}{r_f} \right\}$ and $S_2 = \left\{ {\tilde \phi} \rightarrow - {\tilde \phi} \;,\;  {\tilde \psi} \rightarrow - {\tilde \psi}  \right\}$. 

In the lagrangian (\ref{V2}),  the fact that $\phi$ is a much lighter field than $\psi$ is encoded in the kinetic term, given that $\frac{f_\psi^2}{f_\phi^2} = {\rm O } \left( \alpha^2 \right) \ll 1$. Therefore, in the classical field evolution, the kinetic energy associated to $\dot{\psi}$ can be neglected with respect to that associated to  $\dot{\phi}$ (it would provide a correction to the kinetic term of the same order as those that we have neglected in eq. (\ref{kinetic})). Disregarding the kinetic term of $\psi$ in (\ref{V2}) amounts in treating $\psi$ as a lagrange multiplier. We can then integrate $\psi$ out: 
\begin{equation}
\frac{\partial V}{\partial {\tilde \psi}} = 0 \;\;\; {\rm solved \; by} \;\;\; {\tilde \psi} = {\tilde \psi}_s \left( {\tilde \phi} \right) \;. 
\label{psi-s}
\end{equation} 

This solution provides the light field trajectory in the $2-$field space; this trajectory is stable provided that 
$\frac{\partial^2 V_2}{\partial {\tilde  \psi}^2} > 0 $. When this is the case, the trajectory can support inflation provided that $f_\phi$ is large enough. Inserting the solution (\ref{psi-s}) back into (\ref{V2}) we obtain the one dimensional effective theory 
\begin{equation}
{\cal L} = \frac{f_\phi^2}{2}  \, \left( \partial {\tilde \phi} \right)^2  - \Lambda^4 \, V_1 \left( {\tilde \phi}  \right) \;\;\;,\;\;\; 
V_1 \left( {\tilde \phi} \right) \equiv V_2 \left( {\tilde \phi} ,\, {\tilde \psi}_s \left( {\tilde \phi} \right) \right) \;. 
\label{V1}
\end{equation} 
The $1-$field effective theory (\ref{V1}) has $4$ parameters: the overall potential scale $\Lambda$, the effective axion scale $f_\phi$, and the two dimensionless parameters  $r_f$ and $r_\Lambda$, related to ratios between the axions scales and the potential scales of the original theory, respectively.  Due to the symmetries $S_1$ and $S_2$, the $1-$field effective potential is separately invariant under ${\tilde \phi} \rightarrow - {\tilde \phi}$ and under $\left\{ r_f \rightarrow \frac{1}{r_f} \;,\; r_\Lambda \rightarrow \frac{1}{r_\Lambda} \right\}$. Therefore we can restrict our discussion to $r_f \geq 1$. Specifically, in the main text we focus our attention to $r_f>1$. The limiting case $r_f=1$, which is one of the very few cases for which we could solve (\ref{psi-s}) analytically, is discussed in Appendix \ref{sec:ana}. 
 
We conclude this Section by further discussing the $2-$field potential (\ref{V2}). It is easy to verify that all the critical points of the potential take place when both terms in (\ref{V2}) are at an extremum. If we denote by  ${\cal A}$ and ${\cal B}$ the argument of the two cosines in (\ref{V2}), this happens  when ${\cal A} = \pi \, n$, and ${\cal B} = \pi \, m$, with integer $n$ and $m$. The  $2-$field potential has a minimum when both $m$ and $n$ are even, a maximum when they are both odd, and a saddle point when they are one even and one odd. Moreover, due to the symmetry of the potential under ${\cal A} \rightarrow {\cal A} + 2 \pi  \;,\; {\cal B} \rightarrow {\cal B} + 2 \pi $, we can divide the $\left\{ {\tilde \phi} ,\, {\tilde \psi} \right\}$ plane in an infinite number of equivalent domains.

\begin{figure}
\centerline{
\includegraphics[width=0.35\textwidth,angle=0]{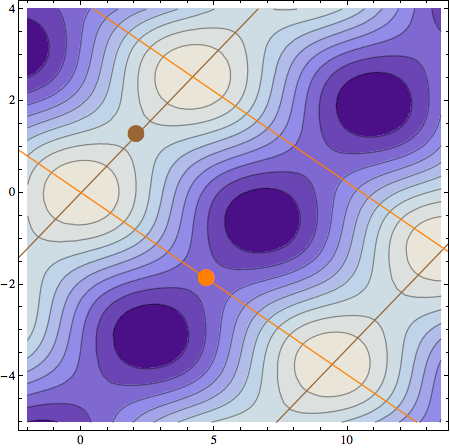}
}
\caption{Contour plot of the potential as a function of the rescaled fields ${\tilde \phi}$ (horizontal direction) and ${\tilde \psi}$ (vertical direction). Due to the different rescaling for the $2-$fields, the potential exhibits  comparable curvature in both directions. However, for $\vert \alpha \vert \ll 1$, the field $\psi$ is significantly heavier than $\phi$, and the evolution proceeds along trajectories where $\frac{\partial V}{\partial {\tilde \psi}} = 0$}
\label{fig:nat-ns-r}
\end{figure}

In Figure \ref{fig:nat-ns-r} we show one such domain, delimited by the lines ${\cal A} =0,\, -2 \pi$ and  ${\cal B} =0,\, 2 \pi$. 
The domain has the origin, with  $\left\{ {\tilde \phi} ,\, {\tilde \psi} \right\}$ coordinates 
\begin{equation}
{\cal O } :\;\;  \left\{ 0,\, 0 \right\} \,,
\end{equation}
in one of its corners (the other three corners are three equivalent minima of the potential). The central part of the domain is occupied by the maximum 
\begin{equation}
{\cal M } :\;\; \pi \left\{ \frac{1}{r_f} + r_f ,\, \frac{r_f^3 \, r_\Lambda - r_f}{1+r_f^4 \, r_\Lambda} \right\} \;. 
\end{equation} 
Finally, each of the four sides contains a saddle point, with equivalent saddle points on opposite sides. We denote by ${\cal S}_A$ and ${\cal S}_B$, respectively, the saddle point on the ${\cal A}=0$ and on the ${\cal B} =0$ line. These points are marked in Figure \ref{fig:nat-ns-r}, and they have coordinates
\begin{equation}
{\cal S}_A = \pi \, \left\{ \frac{1}{r_f} ,\, \frac{r_f^3 \, r_\Lambda}{1+r_f^4 \, r_\Lambda} \right\} \;\;,\;\; 
{\cal S}_B = \pi \, \left\{ r_f ,\, - \frac{r_f}{1+r_f^4 \, r_\Lambda} \right\} \;\;.
\end{equation}  

Without loss of generality, we can restrict the initial conditions for the fields to be in this domain. Moreover - again due to the symmetry properties of the potential - we can restrict the initial conditions to be along the valley that ends on ${\cal O}$, or (when it exists) along a valley that starts from one of the two saddle points ${\cal S}_{A,B}$. Such valleys are shown in Figure  \ref{fig:trajectories}  and are studied in Section \ref{sec:trajectories}. Any other valley in the potential can be mapped to one of these valleys.

\section{Valleys and crests of the $2-$field potential}
\label{sec:trajectories}

\begin{widetext}
\onecolumngrid
\begin{figure}[h!]
\centerline{
\includegraphics[width=0.35\textwidth,angle=0]{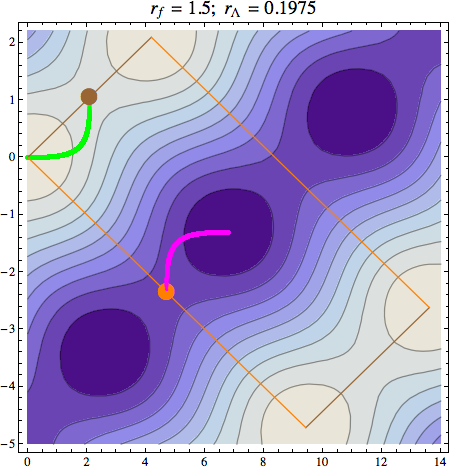}
\includegraphics[width=0.35\textwidth,angle=0]{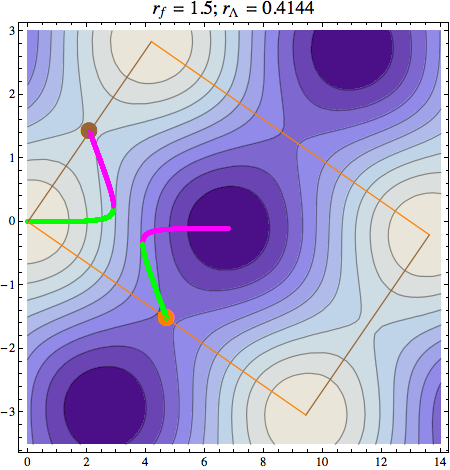}
}
\centerline{
\includegraphics[width=0.35\textwidth,angle=0]{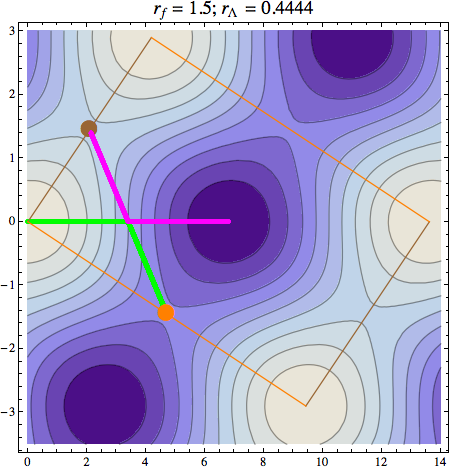}
\includegraphics[width=0.35\textwidth,angle=0]{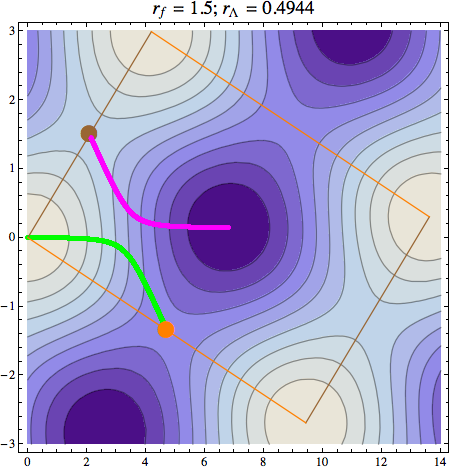}
}
\caption{Contour plot of the potential centered in the domain at the right of the origin. 
The $x-$axis is ${\tilde \phi} = \frac{\phi}{f_\phi}$, while the $y-$axis is  ${\tilde \psi} = \frac{\psi}{f_\psi}$, see eq. (\ref{tilde-fields}). The green (magenta) curves are stable valleys (unstable crests)  within the domain that pass through the origin, the maximum, and / or the saddle points ${\cal S}_{A} ,\, {\cal S}_B$. The plots are for $r_f=1.5$ and for different values of $r_\Lambda$ in the characteristic regions: 
$r_\Lambda \leq \frac{1}{r_f^4}$ (top-left); $\frac{1}{r_f^4} < r_\Lambda < \frac{1}{r_f^2} $ (top-right); $r_\Lambda = \frac{1}{r_f^2}$ (bottom-left); $r_\Lambda > \frac{1}{r_f^2} $ (bottom-right).  
}
\label{fig:trajectories}
\end{figure}
\end{widetext}
\twocolumngrid

We study the valleys and crests  of the potential  ({\ref{V2}); these are lines in the  $\left\{ {\tilde \phi}, \, {\tilde \psi} \right\}$ plane for which $\frac{\partial V_2}{\partial {\tilde \psi}} = 0$. The stable valleys minimize the potential in the $\psi$ direction ($\frac{\partial^2 V_2}{\partial {\tilde \psi}^2} > 0$), while the unstable crests maximize it  ($\frac{\partial^2 V_2}{\partial {\tilde \psi}^2} < 0$). As discussed at the end of the previous Section, we restrict our attention to $r_f > 1$, and to valleys and crests that are (i) inside the domain we have singled out, and (ii) connected either to the origin or to one of the saddle points  ${\cal S}_{A,B}$.  The equipotential lines of Figure \ref{fig:trajectories} show how the valleys and crests extend to other domains, or beyond the maximum ${\cal M}$. However, due to the symmetries of the potential, we do not need to consider these extension in the present discussion. 

In general, we could not solve the condition  $\frac{\partial V_2}{\partial \psi} = 0$ analytically, so that we do not have general analytic expressions for the valleys and crests ${\tilde \psi}_s \left( {\tilde \phi} \right)$. However, it is straightforward to obtain such lines numerically. We summarize the result of our investigation in Figure \ref{fig:trajectories}, where valleys and crests are shown as green and magenta lines, respectively.  The results summarized in the figure  agree with the analytic results that we could obtain in neighborhoods of the extrema of the potential (the origin, the maximum, and  the saddle points), and for the special case $r_\Lambda = \frac{1}{r_f^2}$. This analytic results are reported in Appendix \ref{sec:ana}, and they confirm the validity of what  we present here. The valleys and crests have a qualitatively different behavior according to the relative value of $r_\Lambda$ vs  $r_f$. We discuss these different behaviors starting from small $r_\Lambda$, and discussing how the lines change as $r_\Lambda$ increases. 

For $r_\Lambda \leq \frac{1}{r_f^4}$ (top-left panel of  Figure \ref{fig:trajectories}), a valley extends from the minimum ${\cal O}$ to the saddle point ${\cal S}_A$, and a crest extends from the maximum ${\cal M}$ to the saddle point ${\cal S}_B$. In the interval  $ \frac{1}{r_f^4} < r_\Lambda < \frac{1}{r_f^2}$ (top-right panel) we still find a first curve ${\tilde \psi}_s \left( {\tilde \phi} \right)$ extending from the minimum to ${\cal S}_A$, and a second curve extending from the maximum to ${\cal S}_B$. However both curves are partially a stable valley and partially an unstable crest. For $r_\Lambda = \frac{1}{r_f^2}$ (bottom-left panel), the two curves join each other at a point. Finally, for $r_\Lambda > r_f^2$ (bottom-right panel), a stable valley extends from the origin to ${\cal S}_B$, while an unstable crest extends from the maximum to ${\cal S}_A$. 

At $r_\Lambda= \frac{1}{r_f^2}$, a reconnection between the two ${\tilde \psi}_s \left( {\tilde \phi} \right)$ curves take place. Specifically, for $r_\Lambda < \frac{1}{r_f^2}$ (top two panels of Figure  \ref{fig:trajectories}), a curve  connects the origin with ${\cal S}_A$, and a curve connects the maximum with ${\cal S}_B$. Instead, for  $r_\Lambda > \frac{1}{r_f^2}$ (bottom-right panel of Figure  \ref{fig:trajectories}), the origin is connected with ${\cal S}_B$ and the maximum with ${\cal S}_A$. A ``swap'' between two halves of the two curves takes place at  $r_\Lambda = \frac{1}{r_f^2}$, which is the only value of $r_\Lambda$ for which the two curves touch each other at a point.

\begin{figure}
\centerline{
\includegraphics[width=0.4\textwidth,angle=0]{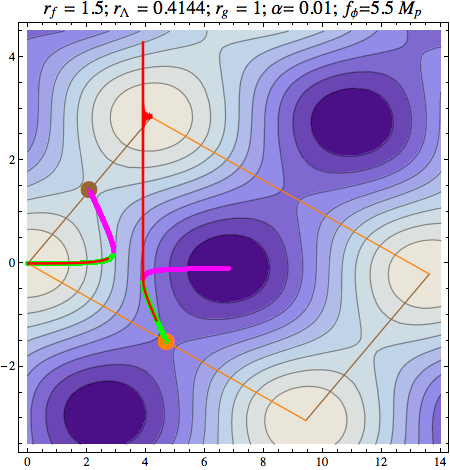}
}
\caption{Contour plot of (\ref{V2}), for parameters in the $\frac{1}{r_f^4} < r_\Lambda < \frac{1}{r_f^2}$ region, together with valleys (green) and crests (magenta). The two red curves are two distinct inflationary trajectories in this model. They are obtained from a numerical evolution of the exact model (\ref{V-start}). Both evolutions shown contain $60$ e-folds of inflation plus a brief transient moment after inflation in which the system reaches the minimum.  
}
\label{fig:traj+evol}
\end{figure}

Provided the effective axion scale $f_\phi$ is sufficiently large, the valleys are inflationary attractors. The valleys shown in Figure \ref{fig:trajectories} are obtained from the potential (\ref{V2}), where ${\rm O } \left( \alpha \right)$ subdominant terms in the exact potential (\ref{V-start}) are disregarded. We performed several numerical evolutions in the exact potential (\ref{V-start}) for $\alpha = \frac{1}{100}$, and we verified that the valleys of Figure  (\ref{fig:trajectories}) describe the inflationary evolution with very good accuracy. Figure \ref{fig:traj+evol} shows two examples of this. We choose parameters in the  $\frac{1}{r_f^4} < r_\Lambda < \frac{1}{r_f^2}$ interval, so that two distinct and inequivalent~\footnote{In the sense that the valleys cannot be mapped into each other by a symmetry of the potential, so they give a different phenomenology.}  valleys exist, one connected to the minimum at the origin, and one disconnected from it. The red curves  in the figure show two inflationary trajectories obtained from the exact potential (\ref{V-start}). These are two different inflationary evolutions for the same model and for the same choice of parameters, but with different initial conditions. As we discuss below, they lead to a different phenomenology. 

The trajectory on the inflationary valley disconnected from the origin is particularly interesting. Inflation ends not because the potential reaches a minimum, but because the inflationary valley ($\frac{\partial V_2}{\partial \psi} > 0$) becomes an unstable crest  ($\frac{\partial V_2}{\partial \psi} < 0)$, so that inflation is terminated by an instability in the heavy direction. We see in the figure that, after leaving the inflationary valley, the system reaches a new $\frac{\partial V_2}{\partial \psi} = 0$ line; the heavy field $\psi$ performs damped oscillations about this valley, and eventually the system reaches the minimum on the top of the figure by evolving in the light $\phi$ direction. This stage of the evolution is very fast. The evolution shown in the figure is characterized by $N=60$ e-folds of inflation along the valley connected to ${\cal S}_B$; the following phase, from the moment the system leaves the valley to when it first reaches the minimum, lasts for $\simeq 2.9$ e-folds. During this second stage, the equation of state oscillates with average  $w_{\rm ave} \simeq -0.16$. This phase should be understood as the beginning of the post-inflationary reheating.

\bigskip \bigskip

\section{$\left\{ n_s-r \right\}$ phenomenology}
\label{sec:pheno}

In this Section we study the CMB phenomenology  of  Aligned Natural Inflation in the $\left\{ n_s-r \right\}$ plane. As discussed in the previous Section, we find two classes of inflationary trajectories in this model: those along valleys connected to a minimum, and those along valleys disconnected from any minimum. In the first case, inflation ends as the fields approach the minimum of the potential; in the second case inflation terminates at the end of the valley, due to an instability in the heavy $\psi$ direction. This second class of solutions exist only for $r_\Lambda$ in the $\left[ \frac{1}{r_f^4} ,\, \frac{1}{r_f^2} \right]$ interval.~\footnote{To be precise, such an evolution can also take place for  $r_\Lambda$ slightly greater than $\frac{1}{r_f^2}$, so that evolutions where inflation ends due to instability in the $\psi$ direction are possible for $\frac{1}{r_f^4} < r_\Lambda < \frac{1}{r_f^2} + \epsilon$, with $\epsilon$ small. The reason for this is that, for $r_\Lambda = \frac{1}{r_f^2}$, one finds $\frac{\partial^2 V_2}{\partial \psi^2} = 0$ at the precise point where the valley connects with the crest shown in the figure, so that the fields do not bend along the valley, but escape from it (ending inflation), and again reach the minimum shown on the top of the figure. This behavior rapidly disappears as $r_\Lambda$ increases slightly above $\frac{1}{r_f^2}$, since $\frac{\partial V_2}{\partial \psi} > 0$ all along the valley in this case. For the parameters used in Figure \ref{fig:traj+evol}, we numerically found that $\epsilon \simeq 0.004$ (while $\frac{1}{r_f^2} \simeq 0.4444$).}

\begin{figure}
\centerline{
\includegraphics[width=0.5\textwidth,angle=0]{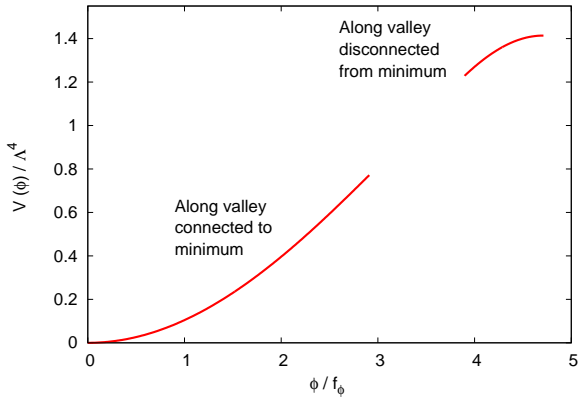}
}
\caption{Potential along the two valleys shown in Figure \ref{fig:traj+evol}. 
}
\label{fig:VcVd}
\end{figure}

Significantly different phenomenological results are obtained in these two cases, and, for this reason, we discuss them in two separate subsections. We anticipate that the evolutions along valleys disconnected from minima lead to a significantly smaller value of the tensor-to-scalar ratio $r$ than the other trajectories, and of what is generically found in axion inflation. The reason for this is that the potential is flatter ( $\equiv$ smaller $\epsilon$ parameter) closer to a saddle point than to a minimum (see  Figure \ref{fig:VcVd}). We stress that the valleys disconnected from the minima are  missed if one rotates the potential at the origin into a heavy $\psi$ and light $\phi$ direction at the origin, and he/she then makes the approximation that all of inflation takes place along the $\psi=0$ line.

We performed several inflationary evolutions for the model, with different values of the  parameters, and we verified that the CMB phenomenology obtained from the $1-$field effective potential (\ref{V1}) is in very good agreement with that obtained from the exact model (\ref{V-start}). In the plots shown below we have obtained the fields evolution from the exact model (\ref{V-start}), and we have then computed the values of $n_s$ and $r$ using the slow-roll relations (\ref{eps-eta}).

\subsection{Inflation on valleys connected to minima}
\label{subsec:pheno-connected}

\begin{figure}
\centerline{
\includegraphics[width=0.4\textwidth,angle=0]{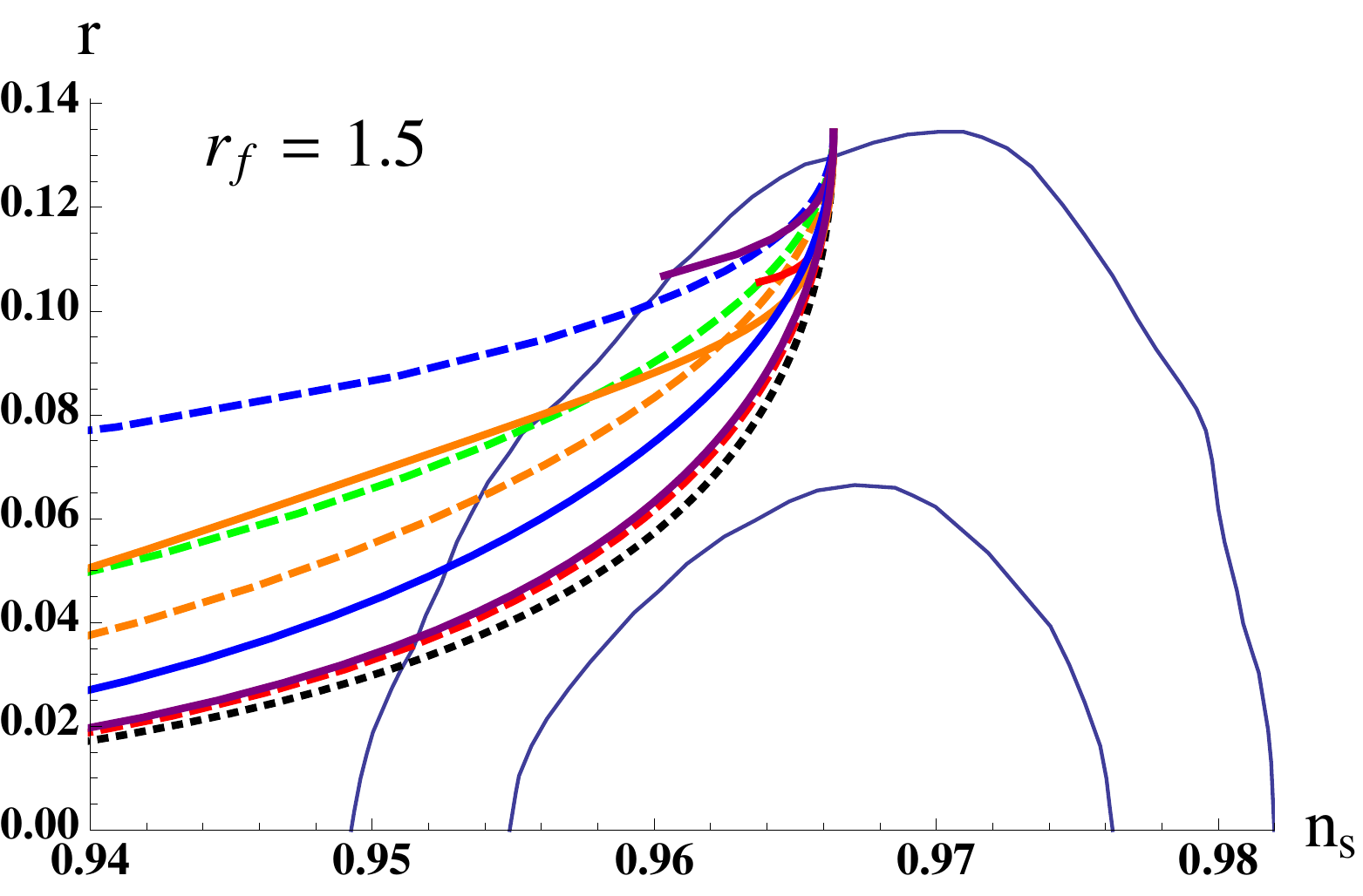}
}
\caption{Predictions of Aligned Natural Inflation (with inflation along a trajectory connected to a minimum) in the $\left\{ n_s - r \right\}$ plane, confronted with the $1\sigma$ and $2 \sigma$ Planck contour lines (specifically, we choose to plot the more conservative red contour lines of Figure 12 of \cite{Ade:2015lrj}).  We fixed $r_f = 1.5$ and varied $r_\Lambda$ as follows: the dashed lines, from bottom to top, are for $r_\Lambda = 0.01, 0.07, 0.1, 0.19$; the solid lines, from top to bottom, are for $r_\Lambda = 0.3, 0.4144, 0.5, 1, 3$. The lowest theoretical curve, drawn as a dotted line, is for Natural Inflation. 
All the theoretical curves are done for $N=60$ e-folds of inflation. 
}
\label{fig:connected-rf1.5}
\end{figure}

For definiteness, we computed the CMB phenomenology for the case in which the axion scales are all comparable to each other, $r_f = 1.5$ and $r_g = 1$. We also fixed the alignment parameter to $\alpha = \frac{1}{100}$, and the number of e-folds of inflation to $N=60$. We then varied $r_\Lambda$ to cover all the different cases shown in Figure \ref{fig:trajectories}. 
The results of our evaluations are shown in Figure \ref{fig:connected-rf1.5}. Each line shown in the Figure corresponds to a given value of $r_\Lambda$ and to varying $f_\phi$. 

All curves have a common end on the top-right part, corresponding to large $f_\phi$. Figure \ref{fig:VcVd} shows the potential along a connected valley as a function of ${\tilde \phi} = \frac{\phi}{f_\phi}$. As $f_\phi$ grows, a given value of ${\tilde \phi}$ produces a longer inflationary expansion. Therefore, if $N$ is fixed to $60$, one needs to decrease the value of $\frac{\phi}{f_\phi}$ as $f_\phi$ increases. At  large $f_\phi$, one is probing only the part of $V \left( \frac{\phi}{f_\phi} \right)$ that is closest to the origin, where a quadratic approximation of the potential suffices. One then recovers the values of massive chaotic inflation, $n_s \simeq 0.967$ and $r \simeq 0.133$. 

As a comparison, the black dotted line in the figure refers to Natural Inflation \cite{Freese:1990rb}. All the four dashed lines in the figure sample the model in the $r_\Lambda < \frac{1}{r_f^4}$ region. We see that the smallest value of $r_\Lambda$ shown gives results very close to Natural Inflation. This is explained by the analytic computation presented in Appendix \ref{sec:ana}, where we show that Aligned Natural Inflation reproduces Natural Inflation in the limit of very  small $r_\Lambda$. 
We then see that increasing $r_\Lambda$ in this interval leads to progressively greater values of $r$. This is particularly true in the left portion of the curves shown in the figure. As we mentioned above, the top-right part of the curves is obtained at relatively large $f_\phi$, while  $f_\phi$ decreases as one moves towards the left of the curves. As $f_\phi$ decreases, ${\tilde \phi}$ needs to start closer and closer to the saddle point. Eq. (\ref{V1-nearSA}) gives the analytic form of the effective $1-$field potential close to this point. We see from this relation that increasing $r_\Lambda$ (in the $r_\Lambda < \frac{1}{r_f^4}$ regime that we are considering here) indeed results in a less flat potential, and so in a greater value of $r$, in agreement with the dashed curves of the figure. 

The two top solid theoretical curves shown in  Figure \ref{fig:connected-rf1.5} sample the $\frac{1}{r_f^4} < r_\Lambda < \frac{1}{r_f^2}$ regime. We recall that in this figure we only consider inflation in the valley connected to the minimum. We see from the top-right panel of Figure \ref{fig:trajectories} that this valley becomes a crest before reaching a saddle point. 
 For this reason, $f_\phi$ cannot be taken arbitrarily small, and, as a consequence, the corresponding curve in the $\left\{ n_s ,\, r \right\}$ plane only extends for a finite interval. The length of this interval decreases as we approach the $r_\Lambda = \frac{1}{r_f^2}$ value. This is due to the fact that also the fraction of the stable portion of the curve from ${\cal S}_A$ to the minimum decreases as $r_\Lambda \rightarrow \frac{1}{r_f^2}$. 

Finally, the three bottom solid theoretical curves shown in  Figure \ref{fig:connected-rf1.5}  sample the $r_\Lambda > r_f^2$ region. 
At relatively small $f_\phi$ (left part of the curves), the evolution starts close to the saddle point ${\cal S}_B$. Eq. (\ref{V1-nearSB}) shows that the potential becomes progressively flatter as $r_\Lambda$ increases in this regime. This explains the behavior of the three bottom solid curves in the figure. We see that the curve corresponding to the highest value of $r_\Lambda$ shown also approaches that of Natural Inflation.  This is explained by the analytic computation presented in Appendix \ref{sec:ana}, where we show that Aligned Natural Inflation reproduces Natural Inflation in the limit of very  large $r_\Lambda$. 

We also performed computations for other values of $r_f$, and we obtained results qualitatively similar to those of Figure 
\ref{fig:connected-rf1.5}. For brevity, we do not show them here.

\subsection{Inflation on valleys disconnected from minima}
\label{subsec:pheno-disconnected}

As we mentioned, in the $\frac{1}{r_f^4} < r_\Lambda < \frac{1}{r_f^2}$ interval, inflation can occur on valleys that are connected to the saddle point ${\cal S}_B$ and that are disconnected from any minimum. Inflation ends because the heavy direction becomes unstable. The tensor-to-scalar ratio in these trajectories is significantly smaller than the one found for the valleys connected to a minimum. This is visible in the two Figures 
\ref{fig:disconnected-rf15} and \ref{fig:disconnected-rf8}, where, respectively, the two cases $r_f = 1.5$ and $r_f = 8$ are studied (such values do not have any particular importance, and they have been chosen just as a representative case of comparable axion scales, or somewhat hierarchical axion scales).

\begin{figure}
\centerline{
\includegraphics[width=0.4\textwidth,angle=0]{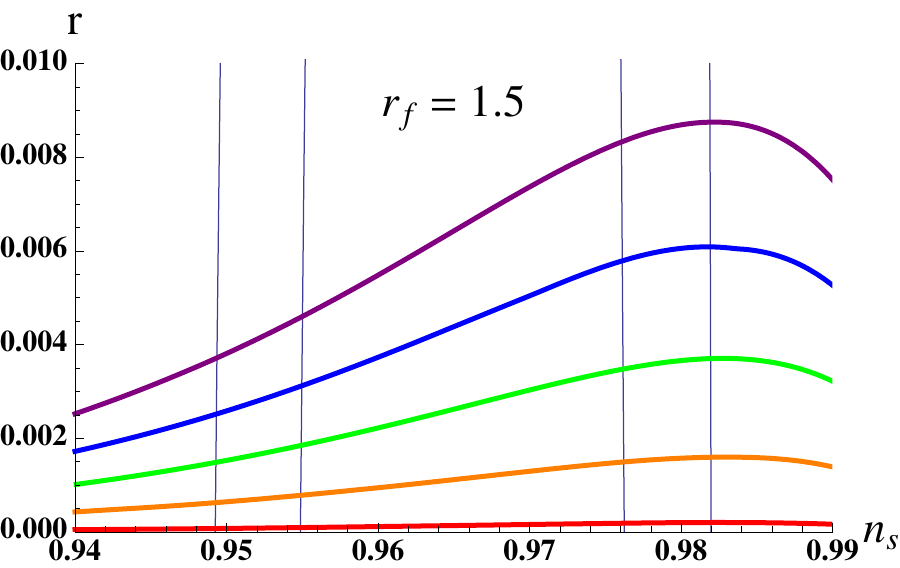}
}
\caption{Predictions of Aligned Natural Inflation (with inflation along a trajectory disconnected from a minimum) in the $\left\{ n_s - r \right\}$ plane, confronted with the $1\sigma$ and $2 \sigma$ Planck contour lines. The theoretical lines have been obtained for $r_f = 1.5$, and for $r_\Lambda$, from bottom to top, equal to $ 0.25, 0.33, 0.38, 0.41, 0.43$. All the theoretical curves are done for $N=60$ e-folds of inflation. 
}
\label{fig:disconnected-rf15}
\end{figure}

\begin{figure}
\centerline{
\includegraphics[width=0.4\textwidth,angle=0]{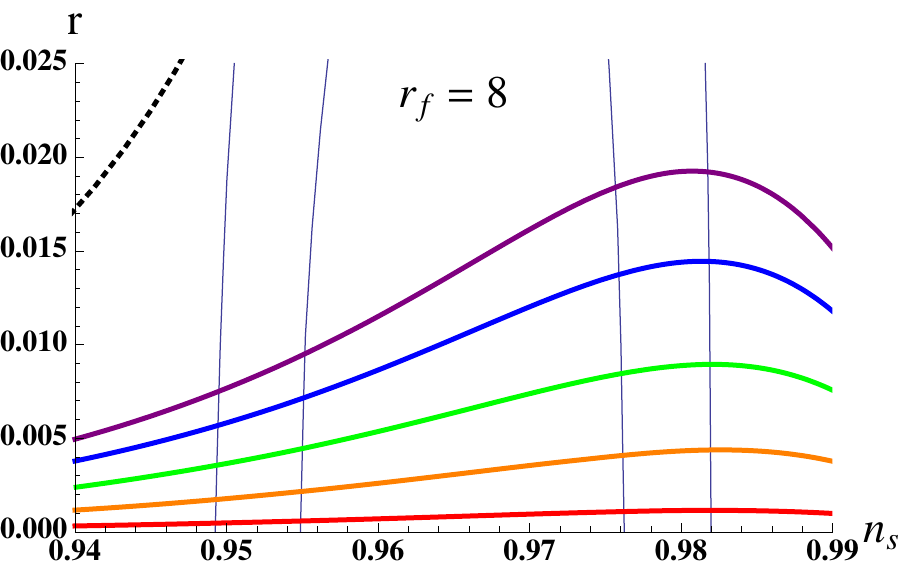}
}
\caption{Predictions of Aligned Natural Inflation (with inflation along a trajectory disconnected from a minimum) in the $\left\{ n_s - r \right\}$ plane, confronted with the $1\sigma$ and $2 \sigma$ Planck contour lines.  The theoretical solid lines have been obtained for $r_f = 8$, and for $r_\Lambda$, from bottom to top, equal to $ 0.005, 0.009, 0.012, 0.014, 0.015$. The dotted line visible in the top-left corner is the theoretical prediction of Natural Inflation. All the theoretical curves are done for $N=60$ e-folds of inflation.}
\label{fig:disconnected-rf8}
\end{figure}

It is possible to reproduce analytically the results shown in these two figures with good accuracy. Most of inflation occurs close to the saddle point, where the $1-$field effective potential reads 
\begin{equation}
V \simeq V_0 \left[ 1 - \left( \frac{\hat \phi}{\hat f} \right)^2 \right] \;\;,\;\; 0 \leq {\hat \phi} \leq {\hat \phi}_0 \;. 
\end{equation}
In this equation, we have defined  (cf. eq. (\ref{V1-nearSB})), 
\begin{equation}
{\hat \phi} = \pi r_f f_\phi - \phi \;\;,\;\; {\hat f} = \frac{2 f_\phi}{r_f} \, \sqrt{\frac{r_f^4 r_\Lambda - 1}{r_\Lambda}} \;\;, 
\end{equation}
while ${\hat \phi}_0$ denotes the value of ${\hat \phi}$ at which inflation ends (due to the fact that the orthogonal direction $\psi$ becomes tachyonic at that point). This is a top-hat potential with maximum on the saddle point ${\cal S}_B$ 
(of the $2-$field potential) located at ${\hat \phi} = 0$. The slow roll parameters on this potential are 
\begin{equation}
\epsilon \simeq \frac{2 M_p^2 {\hat \phi}^2}{{\hat f}^4} \;\;,\;\; 
\eta \simeq - \frac{2 M_p^2}{{\hat f}^2} \;\;,\;\; {\hat \phi} \ll {\hat f} \;. 
\end{equation} 
Integrating the slow roll relation $\dot{\hat \phi} \simeq \sqrt{2 \epsilon} H M_p$, one  obtains the value of ${\hat \phi}$ at $N$ e-folds before the end of inflation: 
\begin{equation}
{\hat \phi}_N \simeq {\hat \phi}_0 \, {\rm e}^{- \frac{2 M_p^2 N}{{\hat f}^2} } \;\;. 
\end{equation} 
and finally, from the slow roll relations $r \simeq 16 \epsilon \;,\; n_s - 1 \simeq 2 \eta - 6 \epsilon \simeq 2 \eta$, one obtains the functional relation 
\begin{equation}
r \simeq 8 \left( 1 - n_s \right) \, \frac{{\hat \phi}_0^2}{{\hat f}^2 } \, {\rm e}^{- N \left( 1 - n_s \right)} \;\;. 
\label{ns-s-disconnected}
\end{equation} 

Each theoretical curve in Figures \ref{fig:disconnected-rf15} and \ref{fig:disconnected-rf8} is obtained for fixed $r_f$ and $r_\Lambda$, and for varying $f_\phi$. The value of ${\hat \phi}_0$ can be obtained from imposing $\frac{\partial^2 V_2}{\partial \psi^2} =0$ along the inflationary trajectory, where the potential (\ref{V2}) is used. We see that any fixed value of $r_f$ and $r_\Lambda$, results in a given value of $ \frac{{\hat \phi}_0^2}{{\hat f}^2 } $.  Therefore, each curve is characterized by the same value~\footnote{This is confirmed with very good accuracy by the numerical evolutions with the exact potential  (\ref{V-start}).} of $\frac{{\hat \phi}_0^2}{{\hat f}^2 } $, so that (\ref{ns-s-disconnected}) gives the prediction for the shape of the theoretical curves, up to an overall rescaling. A comparison with  the exact numerical results shown in Figures \ref{fig:disconnected-rf15} and \ref{fig:disconnected-rf8} shows that (\ref{ns-s-disconnected}) -  with ${\hat \phi}_0$ obtained from a numerical evolution with the exact $2-$field potential -  is accurate up to a few percent level discrepancy for $r_\Lambda \simeq \frac{1}{r_f^2}$, while the discrepancy increases to $\sim 10- 50\%$ level for  $r_\Lambda \rightarrow \frac{1}{r_f^4}$. 

We can have a fully analytic relation by estimating the value ${\hat \phi}_0$ at which inflation vanishes. Eq. (\ref{SB-traj}) is an analytic expression for the inflationary trajectory in the neighborhood of ${\cal S}_B$. We insert this expression in $\frac{\partial^2 V_2}{\partial \psi^2}$, expand to quadratic order in $\delta {\tilde \phi}$, and set this expression to zero. This provides a simple estimate for the value of ${\hat \phi}$ at which the heavy direction becomes tachyonic: 
\begin{equation}
\frac{{\hat \phi}_0}{\hat f} \simeq \frac{r_f^4 r_\Lambda - 1 }{\sqrt{2} \sqrt{1-r_\Lambda} \, r_f^2} \,. 
\label{end-analytic}
\end{equation} 
We stress that this value cannot be extremely accurate, as it is obtained by expanding the potential close to ${\cal S}_B$, and not close to the where the instability occurs. Nonetheless, it provides a reasonable estimate. By comparing with the exact numerical evolutions used for  Figures \ref{fig:disconnected-rf15} and \ref{fig:disconnected-rf8}, we find that the estimate (\ref{end-analytic}) is accurate up to $\sim 10\%$ discrepancy for  $r_\Lambda \simeq \frac{1}{r_f^2}$, while the discrepancy increases to $\sim 20- 30\%$ level as  $r_\Lambda \rightarrow \frac{1}{r_f^4}$. Combining (\ref{ns-s-disconnected}) and (\ref{end-analytic}), one obtains 
\begin{equation}
r \simeq \frac{4}{r_f^4} \,\,   \frac{\left( r_f^4 r_\Lambda - 1 \right)^2}{ 1-r_\Lambda } \,   \left( 1 - n_s \right)  \, {\rm e}^{- N \left( 1 - n_s \right)} \;\;. 
\label{ns-s-analytic}
\end{equation} 

Despite not perfectly accurate, eq. (\ref{ns-s-analytic}) can be used to understand the behavior of $r$ with the parameters of the potential. For instance, it shows that,  at any fixed $r_f$, the value of $r$ grows with growing $r_\Lambda$, in agreement with what is seen in Figures  \ref{fig:disconnected-rf15} and \ref{fig:disconnected-rf8}. It also shows that $r$ in general grows with $r_f$ (the derivative with respect to $r_f$ of the right hand side of (\ref{ns-s-analytic}), evaluated for the $r_\Lambda > \frac{1}{r_f^4}$, is positive). This is also confirmed by the comparison between Figure  \ref{fig:disconnected-rf15} and Figure  \ref{fig:disconnected-rf8}.

\section{Coupling to vector fields}

\label{sec:phiFF}

The couplings of an axion to ``matter'' fields are highly constrained by the shift symmetry. The leading operators  that give the coupling to gauge and fermion fields are, respectively, $\frac{c_A}{4 \, f } \, \phi \, F \, {\tilde F}$ and $\frac{c_\psi}{f} \, \partial_\mu \phi \, {\bar \psi} \gamma_5 \gamma^\mu \psi$, where $f$ is the axion scale, and where $c_A$ and $c_\psi$ are model dependent dimensionless coefficients that, in the spirit of an effective field theory, are naturally expected to be ${\rm O} \left( 1 \right)$.

The coupling to gauge fields typically controls the perturbative decay of an axion inflaton after inflation, as the decay into fermions is helicity suppressed  \cite{Pajer:2013fsa}. Moreover, this coupling also leads to interesting gauge field production already during inflation.~\footnote{These considerations are valid for any gauge field whose quanta are light during inflation. Although explicit computations in the literature assume U(1) fields, the results can be easily extended to the non-abelian case. See \cite{Pajer:2013fsa} and \cite{Barnaby:2011vw}  for the details of the computations that we discuss in this section.}   Specifically, one gauge field helicity becomes tachyonic in presence of this operator \cite{Anber:2006xt}, and the corresponding gauge quanta are non-perturbatively produced (this effect vanishes both in  the UV regime and IR regime, so the production is ``self-regulated'', and does not diverge). The amount of produced quanta~\footnote{In this discussion we consider the production of gauge fields from the inflaton during inflation. See \cite{Barnaby:2011qe} for a discussion and \cite{Adshead:2015pva} for a more detailed study of the production at reheating.}  is exponentially sensitive to the parameter \cite{Anber:2006xt} 
$
\xi \equiv \frac{c_A \, \vert \dot{\phi} \vert}{2 \, f \, H} \,, 
$
in the $\xi > 1$ regime, while the production is negligible at $\xi < 1$ (for definiteness, we assume $\xi$ to be positive in this discussion. Changing sign of $\xi$ simply changes the gauge helicity state that is produced, and so it does not change the phenomenological constraints). These gauge quanta inverse decay to produce inflaton perturbations, that are highly non-gaussian \cite{Barnaby:2010vf} and that have a blue spectrum that can lead to  too many primordial black holes  \cite{Linde:2012bt}. They can also produce gravity waves  \cite{Barnaby:2011vw} that violate parity \cite{Sorbo:2011rz}, and that also have a blue spectrum, growing to a level that can be tested already at the next generation interferometer experiments \cite{Cook:2011hg,Barnaby:2011qe,Crowder:2012ik}. All these effects constrain $\xi \la {\rm O } \left( 1 \right)$. The estimate of  \cite{Linde:2012bt} suggests that 
the limit from primordial black holes, that forces $\xi \la 1.5$, is the strongest one, although the precise limit is sensitive to the evolution of the  scalar perturbations in a highly nonlinear regime, which may not be fully under computational control  \cite{Linde:2012bt}. The other limits are more robust, and constrain $\xi \la 2 - 2.5$  \cite{Pajer:2013fsa}.~\footnote{The quantity $\xi$ typically increases during inflation. The limits quoted here refer to the value assumed by $\xi$ when the large CMB scales left the horizon.} Using the slow roll relation $\vert \dot{\phi} \vert \simeq \sqrt{2 \epsilon} \, H \, M_p$, we can rewrite 
\begin{equation}
\xi = \frac{c_A \, M_p}{f} \, \sqrt{\frac{\epsilon}{2}} \;. 
\end{equation}
and so we see that the regime $\xi \ga {\rm O } \left( 1 \right)$ is typically reached when $\frac{f}{c_A} \la 10^{-2} M_p$  \cite{Barnaby:2010vf}. 

Such effects may help in phenomenologically distinguishing the single field Natural Inflation from the Aligned Inflation mechanism. Indeed, as we now discuss, they are negligible in the former case, while they may be detectable in the latter case. To see this, consider the operators
\begin{eqnarray}
\Delta {\cal L}_{\rm Natural} &=& - \frac{1}{4} F^2- \frac{1}{4 } \, c \, \frac{\phi}{f} \, F \, {\tilde F} \;, \nonumber\\
\Delta {\cal L}_{\rm Aligned} &=& - \frac{1}{4} F^2- \frac{1}{4 } \, \left( c_\theta \, \frac{\theta}{f}  + c_\rho \, \frac{\rho}{f} \right)  F \,  {\tilde F} \;,
\label{phiFFt}
\end{eqnarray}
where $f$ in the second line is the axion scale introduced in (\ref{def-f-rf}). The parameters $c, c_\theta, c_\rho$ are model-dependent dimensionless coefficients  (for instance, such a coupling can arise from integrating out heavy fermions that couple both to the axion and to the gauge field), which may naturally be expected to be of order one. In the aligned inflation case, we use (\ref{rotation}) to rotate to the base of light ($\phi$) and heavy field ($\psi$) during inflation, and we arrive to the coupling 
\begin{equation}
\Delta {\cal L}_{\rm aligned}=-\frac{1}{4}F^2-\frac{1}{4}\left(  \frac{C_{\phi}\,\phi}{\alpha f_{\phi}}+\frac{C_{\psi}\,\psi}{f_{\psi}}  \right)F{\tilde F}
\label{phiFFt2}
\end{equation}
where the scales $f_\phi$ and $f_\psi$ have been defined in (\ref{tilde-fields}), while, again, the parameters $C_\phi$ and $C_\psi$ are model-dependent dimensionless coefficients  which may naturally be expected to be of order one.~\footnote{These coefficients are related to those entering in (\ref{phiFFt}) by  $C_{\phi}=\frac{c_{\theta}-c_{\rho}\,r_g}{2}  $ and $ C_{\psi}=\frac{\left(c_\rho + r_g \, c_\theta \right) r_g}{1+r_g^2} $, up to ${\rm O } \left( \alpha \right)$ corrections.} 

As we have discussed after eq. (\ref{tilde-fields}), the rescaled fields ${\tilde \phi} = \frac{\phi}{f_\phi}$ and  ${\tilde \psi} = \frac{\psi}{f_\psi}$ exhibit a similar excursion during inflation. However, the original field $\phi$ moves much more than $\psi$. This is encoded in the $\frac{1}{\alpha}$ factor entering in (\ref{phiFFt2}): due to the greater excursion, the  light field $\phi$ is coupled to the gauge field much more strongly than $\psi$. So, the coupling of the inflaton to the gauge field in Natural Inflation and Aligned Natural Inflation is given, respectively, by 
\begin{equation}
\xi_{\rm natural} =  c \, \frac{M_p}{f} \,  \sqrt{\frac{\epsilon}{2}}  \;\;,\;\; 
\xi_{\rm aligned} =  \, \vert C_\phi \vert \, \frac{M_p}{\alpha \, f_\phi} \,  \sqrt{\frac{\epsilon}{2}}  \;.
\end{equation} 
We therefore see that, in Natural Inflation, a value $c \ga {\rm O } \left( 10^2 - 10^3 \right)$ is necessary to produce a visible effect, in contrast to the natural  $c \la {\rm O } \left( 1 \right)$ expectation. Also in Aligned Natural Inflation, a value 
$\frac{C_\phi}{a} \ga {\rm O } \left( 10^2 - 10^3 \right)$ is required to have a visible effect. However, due to the $\frac{1}{\alpha} \gg 1$ enhancement, this can be possible even if  $C_\phi \la {\rm O } \left( 1 \right)$.

\begin{figure}
\includegraphics[width=0.45\textwidth,angle=0]{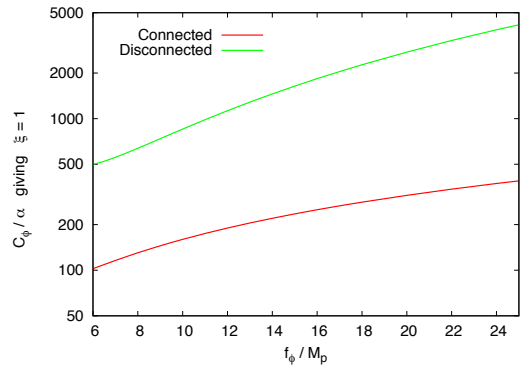}
\caption{The quantity $\frac{C_\phi}{\alpha}$ is a dimensionless coefficient controlling the coupling of the inflaton in Natural Aligned Inflation to a light gauge field (defined in eq. (\ref{phiFFt2})). The curves show that value of this coefficient leading to observable gauge quanta production during inflation, for the special choice of parameters $r_f = 1.5$ and $r_\Lambda = 0.4$, and for an inflationary trajectory  along a valley that is connected to / disconnected from a minimum of the potential.}
\label{fig:FFt}
\end{figure}

In Figure \ref{fig:FFt} we show the required value of $\frac{C_\phi}{a} $ to produce $\xi =1$ in Aligned Natural Inflation, for the parameter choice $r_f =1.5 ,\, r_\Lambda = 0.43$.~\footnote{Figure \ref{fig:FFt}  has been obtained from the exact value of $\dot{\phi}$ in numerical evolutions of the exact $2-$field model (\ref{V-start}), with $\alpha = 0.01$ and $r_g=1$. However, the same result can also be obtained from the effective $1-$field model (\ref{V1}), where the parameters $\alpha$ and $r_g$ do not enter (one only requires that $\alpha \ll 1$).} We observe that a stronger coupling is required for trajectories along valleys disconnected from minima of the potential, since the inflaton potential is flatter along those valleys (see the discussion at the beginning of Section \ref{sec:pheno}). 

To summarize, in Natural Inflaton the request of a sufficiently flat potential, namely $f > M_p$, imposes a strong limit on the coupling strength $\frac{1}{f} \times {\rm O } \left( 1 \right)$ of the inflaton to gauge fields, which precludes an interesting production of gauge quanta during inflation. On the contrary, if the aligned mechanism is employed, one can start from significantly smaller axion decay constants, and hence a significantly stronger coupling to gauge fields, and still obtain a flat inflaton potential, $f_\phi = {\rm O } \left( \frac{f}{\alpha} \right) > M_p$. The fact that none of the effects associated to the gauge field production has been yet observed translates into limits of the coupling of the axions to any light gauge field, and provides useful constrains for model building  \cite{Barnaby:2011qe}.

\section{Conclusions}
\label{sec:conclusion}

The alignment mechanism of \cite{Kim:2004rp} has received considerable attention in the recent past. Despite of this, a detailed study of the inflationary trajectories in this model, and of the associated phenomenology, was still missing. The main goal of this work was to perform such a study. We were particularly interested in understanding whether the model can be phenomenologically distinguished from single field Natural Inflation. The prediction from Natural Inflation
 are outside the current $1 \sigma$ CMB contours, and the model will be fully explored once the sensitivity to $r \sim 0.03$ is reached~\cite{Munoz:2014eqa}, see for instance Figure \ref{fig:connected-rf1.5}.~\footnote{In our figures,  we fixed the number of e-folds of visible inflation to $N_{\rm CMB} =60$. Such a value is obtained for a quick reheating after Natural Inflation. Slower reheating (under the conventional assumption that the equation of state  at reheating is smaller than that of radiation \cite{Podolsky:2005bw}, $w_{\rm reh} \leq \frac{1}{3}$)  implies that the large scale CMB perturbations were produced closer to the end of inflation ($N_{\rm CMB} < 60$), when the potential was less flat. This gives a larger value of $r$ than the one shown in Figure \ref{fig:connected-rf1.5}.} A number of current and near future CMB experiments have claimed $r = {\rm O } \left( 10^{-2} \right)$ sensitivity, and the current discussion of Stage 4 experiments aims to reach $r = {\rm O } \left( 10^{-3} \right)$ sensitivity \cite{Abazajian:2013vfg}. It is therefore of particular interest to study whether Aligned Natural Inflation can provide values of $n_s - r$  below the dotted line shown in  Figure \ref{fig:connected-rf1.5}, as this region cannot be covered by single field Natural Inflation. 

The most interesting result emerged from our analysis is the existence of inflationary trajectories near saddle points of the   potential, and disconnected from minima of the potential (in the sense that these trajectories end far away from minima, because the heavy field perpendicular to them becomes unstable). As studied in Section \ref{sec:pheno} these trajectories lead to much smaller values of $r$ than those obtained in  Natural Inflation.  The $n_s - r$ prediction for the evolutions along these trajectories is given by eq.  (\ref{ns-s-analytic}), which is a fully analytic expression  in terms of the parameters in the model and of the number of e-folds. As we discussed, this relation is only an approximate one, but it is reasonably accurate, and it allows to understand  the dependence of the phenomenological predictions on the parameters of the model. 

A second possible way to distinguish Aligned Natural Inflation from single field Natural Inflation originates from the fact that a stronger coupling of the inflaton to other fields should be expected in the former, given that the starting axion scales are sub-Planckian. A moderate level of alignment, $\alpha = {\rm O } \left( 10^{-2} \right)$, is associated to ${\rm O } \left( 10^{16} \, {\rm GeV} \right)$ axion scales, where the coupling of the axions to any massless or light ($m < H$) gauge field can 
lead to significant gauge field production during inflation, with several phenomenological consequences  \cite{Pajer:2013fsa}. We reviewed this in Section \ref{sec:phiFF}. 

Finally, we  hope that the conclusions learnt from the present study can be of relevance also for other multi-field models of axion inflation. Recently, there has been considerable interest of inflation in the landscape of string theory, and on the relevance of critical points in multi-field inflationary potentials, see for instance \cite{Aazami:2005jf,Frazer:2011tg,Agarwal:2011wm,Battefeld:2012qx,Battefeld:2013xwa,Marsh:2013qca}. We have shown that the presence of critical points in one of the $2-$field extensions of Natural Inflation can strongly impact the phenomenological predictions of that model, and this may be the case also for other multi-field extensions that we have mentioned in this work. It may be interesting to perform a systematic study of trajectories, and explore the relevance of saddle points, also for such models.

\vskip.25cm
\noindent{\bf Acknowledgements:} 
The work of M.P. is partially supported from the DOE grant DE-SC0011842  at the University of Minnesota. The work of C.U. was partially supported by a summer grant from the graduate program of the School of Astronomy and Physics of the University of Minnesota.

\appendix 

\section{Analytic properties of the valleys and crests}
\label{sec:ana}

In this Appendix, we denote as ``curve'' a line  ${\tilde \psi}_s \left( {\tilde \phi} \right)$ along which $\frac{\partial V_2}{\partial {\tilde \psi}} = 0$. Namely
\begin{equation}
\sin \left( - \frac{r_f^3 \, r_\Lambda \, {\tilde \phi}}{1+ r_f^4 \, r_\Lambda} + \frac{{\tilde \psi}_s}{r_f} \right) + 
r_f^2 \, r_\Lambda \sin \left(  \frac{r_f  \, {\tilde \phi}}{1+ r_f^4 \, r_\Lambda} + r_f \, {\tilde \psi_s} \right) = 0 \,. 
\label{minimumPsi}
\end{equation} 
These curves are stable valleys if $\frac{\partial^2 V_2}{\partial {\tilde \psi^2}} \vert_{{\tilde \psi} = {\tilde \psi}_s} > 0$, in which case they can support inflation provided that the effective axion scale $f_\phi$ is sufficiently large. They are instead unstable crests if  $\frac{\partial^2 V_2}{\partial {\tilde \psi^2}} \vert_{{\tilde \psi} = {\tilde \psi}_s} < 0$. 

Inserting the solution ${\tilde \psi}_s \left( {\tilde \phi} \right)$ of (\ref{minimumPsi}) into the $2-$field potential $V_2$ provides the effective $1-$field potential, as formally described in eq. (\ref{V1}). In general, eq. (\ref{minimumPsi}) can be solved numerically. Here we discuss the cases for which we could obtain an analytic solution.

\subsection{Critical points} 

The condition (\ref{minimumPsi}) is obviously satisfied at the critical points ${\cal O},\, {\cal M} ,\, {\cal S}_A ,\, {\cal S}_B$ 
listed in Section  \ref{sec:nkp-review}. We can therefore solve it analytically for small displacements about these points. 
Next to the origin we obtain 
\begin{equation}
{\tilde \psi}_s \simeq - \frac{r_f^6 \, r_\Lambda}{6} \, \frac{r_f^4 \, r_\Lambda^2-1}{\left( r_f^4 \, r_\Lambda +1 \right)^4} \: {\tilde \phi}^3 \,. 
\label{O-traj}
\end{equation} 

By computing  $\frac{\partial^2 V_2}{\partial {\tilde \psi^2}}$, and imposing (\ref{O-traj}), we can verify that (\ref{O-traj}) is  a stable valley  (which is obvious, given that the origin is a minimum of $V_2$). We observe that, as the valley reaches the origin, ${\tilde \psi}_s$ vanishes faster than ${\tilde \phi}$. This is due to the fact that, as we discussed after performing  the rotation (\ref{rotation}), the $2-$fields $\phi$ and $\psi$ are mass eigenvectors near the origin, so that the light field valley must proceed along $\psi =0$ at this point. We also see from (\ref{O-traj}) that, on the immediate right of the origin, the valley bends to positive ${\tilde \psi}$  for $r_\Lambda < \frac{1}{r_f^2}$ and to  negative ${\tilde \psi}$  for  $r_\Lambda > \frac{1}{r_f^2}$. This can be seen, respectively, in the top two panels of Figure \ref{fig:trajectories} and in the bottom right panel. 

Being of ${\rm O } \left( \phi^3 \right)$, the field $\psi$ does not contribute to the quadratic potential near the origin, where we have 
\begin{equation}
 V_{1,{\rm near \, {\cal O}}} \simeq \frac{1}{2} \, \frac{r_f^2 \, r_\Lambda}{\left( 1 + r_\Lambda \right) \left( 1 + r_f^4 \, r_\Lambda \right)} \, {\tilde \phi}^2 \,. 
\end{equation} 

Near the maximum, namely for ${\tilde \phi} = \pi \left( \frac{1}{r_f} + r_f \right) + \delta {\tilde \phi}$ and 
${\tilde \psi}_s = \pi \, \frac{r_f^3 \, r_\Lambda - r_f}{1+r_f^4 \, r_\Lambda} + \delta {\tilde \psi}$, the condition (\ref{minimumPsi}) is solved by 
\begin{equation}
\delta {\tilde \psi} \simeq - \frac{r_f^6 \, r_\Lambda}{6} \, \frac{r_f^4 \, r_\Lambda^2-1}{\left( r_f^4 \, r_\Lambda +1 \right)^4} \: \delta {\tilde \phi}^3 \,.
\end{equation} 
This curve is unstable near ${\cal M}$. To the left of the maximum, this curve  bends towards negative ${\tilde \psi}$ for $r_\Lambda < \frac{1}{r_f^2}$ and towards positive ${\tilde \psi}$ for $r_\Lambda > \frac{1}{r_f^2}$. This is the opposite of the curve connected to the origin. As Figure \ref{fig:trajectories} shows, when one curve connects the minimum with the  saddle point ${\cal S}_A$ at positive ${\tilde \psi}$, the other curve connects the maximum  with the saddle point  ${\cal S}_B$ at negative ${\tilde \psi}$, and viceversa. 

Near the saddle point ${\cal S}_A$, namely for ${\tilde \phi} = \frac{\pi}{r_f} + \delta {\tilde \phi}$ and ${\tilde \psi}_s = \frac{\pi r_f^3 r_\Lambda}{1+r_f^4 r_\Lambda} +  \delta {\tilde \psi}$,  the condition (\ref{minimumPsi}) is solved by 
\begin{equation}
\delta {\tilde \psi} \simeq  \frac{2 \, r_f^4  \,  r_\Lambda}{1- r_f^8 \, r_\Lambda^2}  \: \delta {\tilde \phi} \,. 
\label{SA-traj}
\end{equation} 
Expanding  $\frac{\partial^2 V_2}{\partial {\tilde \psi^2}}$ along (\ref{SA-traj}), we find that this curve is stable near ${\cal S}_A$ for $r_\Lambda < \frac{1}{r_f^4}$, which is the case shown in the top-left panel of Figure  \ref{fig:trajectories}. Along the curve, and close to ${\cal S}_A$, the potential reads 
\begin{equation}
 V_{1,{\rm near \, {\cal S}_A}}  \simeq   \frac{2 \, r_\Lambda}{1+r_\Lambda} \left[ 1 - \frac{1}{4} \, \frac{r_f^2}{ 1 - r_f^4 \, r_\Lambda }  \, \delta {\tilde \phi}^2 \right] \,. 
\label{V1-nearSA}
\end{equation}

 Finally, near the saddle point ${\cal S}_B$, namely for  ${\tilde \phi} = \pi\, r_f + \delta {\tilde \phi}$ and ${\tilde \psi}_s = - \frac{\pi r_f }{1+r_f^4 r_\Lambda} +  \delta {\tilde \psi}$,   the condition (\ref{minimumPsi}) is solved by 
\begin{equation}
\delta {\tilde \psi} \simeq  \frac{2 \, r_f^4  \,  r_\Lambda}{1- r_f^8 \, r_\Lambda^2}  \: \delta {\tilde \phi} \,. 
\label{SB-traj}
\end{equation} 
Expanding  $\frac{\partial^2 V_2}{\partial {\tilde \psi^2}}$ along (\ref{SB-traj}), we find that this curve is stable near ${\cal S}_B$ for $r_\Lambda > \frac{1}{r_f^4}$, which is the case for  the top-right panel and the two bottom panels of Figure  \ref{fig:trajectories}. Along the trajectory, and close to ${\cal S}_B$, the potential reads 
\begin{equation}
 V_{1,{\rm near \, {\cal S}_B}}  \simeq   \frac{2  }{1+r_\Lambda} \left[ 1 - \frac{1}{4} \, \frac{r_f^2 \, r_\Lambda}{  r_f^4 \, r_\Lambda -1 }  \, \delta {\tilde \phi}^2 \right] \,. 
 \label{V1-nearSB}
\end{equation}

\subsection{$r_\Lambda = \frac{1}{r_f^2}$} 

As discussed in the main text, at the interesting value of $\,r_\Lambda= \frac{1}{r_f^2}$, a reconnection between the two ${\tilde \psi}_s \left( {\tilde \phi} \right)$ curves takes place. For  $r_\Lambda= \frac{1}{r_f^2}$, the field ${\tilde \phi}$ enters in the two arguments of the two sine functions in (\ref{minimumPsi}) multiplied by the same coefficient $\frac{r_f}{1+r_f^2}$. We could then solve (\ref{minimumPsi}) analytically, and find that the condition  $\frac{\partial V_2}{\partial {\tilde \psi}}$ is satisfied by two straight lines: one joining the minimum ${\cal O}$ with the maximum ${\cal M}$, and one joining the two saddle points ${\cal S}_A$ and ${\cal S}_B$. The two lines intersect at the point ${\cal C}$, of coordinates
\begin{equation}
{\cal C } :\;\; \frac{\pi}{2} \left\{ \frac{1}{r_f} + r_f ,\, 0  \right\} \;. 
\end{equation} 

One can show analytically that only parts of the two curves are stable ($\frac{\partial^2 V_2}{\partial {\tilde \psi^2}} > 0$): specifically,  the segment between ${\cal C}$ and the minimum ${\cal O}$, and  the segment between ${\cal S}_B$ and ${\cal C}$. The other parts of the trajectories are unstable. 

Along the stable ${\cal C}-{\cal O}$ segment, the potential reads 
\begin{equation}
V_1 =  1 - \cos \left( \frac{r_f}{1+r_f^2} \, {\tilde \phi} \right)   \;. 
\end{equation}
and the argument of the cosine ranges from $\frac{\pi}{2}$ to $0$ as one moves from ${\cal C}$ to ${\cal O}$ along the trajectory.   

Along the stable ${\cal S}_B-{\cal C}$ segment, the potential reads 
\begin{equation}
V_1 =  1 + \frac{r_f^2-1}{r_f^2+1} \, \sin \left[ \frac{r_f}{r_f^2-1} \, \left(  {\tilde \phi} - \frac{\pi \left( 1 + r_f^2 \right)}{2 r_f}  \right) \right]   \;, 
\end{equation} 
and the argument of the sine ranges from $\frac{\pi}{2}$ to $0$ as one moves from ${\cal S}_B$ to ${\cal C}$ along the curve.

\subsection{$r_f =1$} 

We recall that the potential (\ref{V2}) is invariant under the simultaneous  exchanges $r_f \rightarrow \frac{1}{r_f}$ and 
$r_\Lambda \rightarrow \frac{1}{r_\Lambda}$. For this reason, we do not need to discuss the case $r_f<1$. In the main text we studied the case $r_f>1$. Here we discuss the $r_f=1$ case. The main difference with the discussion of the main text is that now the regime  $\frac{1}{r_f^4} < r_\Lambda < \frac{1}{r_f^2}$  is absent. 

For $r_f=1$, the field ${\tilde \psi}$ enters with the same coefficient in the two potential terms. Thanks to this, we can write
\begin{eqnarray}
V_2 &=& 1 - \frac{A \left( {\tilde \phi} \right) \cos {\tilde \psi} + B \left( {\tilde \phi} \right) \sin {\tilde \psi}}{1+r_\Lambda} \;, \nonumber\\ 
\frac{\partial V_2}{\partial {\tilde \psi}} &=& \frac{A \left( {\tilde \phi} \right) \sin {\tilde \psi} - B \left( {\tilde \phi} \right) \cos {\tilde \psi}}{1+r_\Lambda} \;\;,\;\; 
\frac{\partial^2 V_2}{\partial {\tilde \psi}^2} = 1 - V_2 \;, \nonumber\\ 
\label{rf1-AB}
\end{eqnarray} 
where
\begin{eqnarray}
A \left( {\tilde \phi} \right) &=& \cos \frac{r_\Lambda \, {\tilde \phi}}{1+r_\Lambda} + r_\Lambda \, \cos \frac{\tilde \phi}{1+r_\Lambda}  \;, \nonumber\\ 
B \left( {\tilde \phi} \right) &=& \sin \frac{r_\Lambda \, {\tilde \phi}}{1+r_\Lambda} - r_\Lambda \, \sin \frac{\tilde \phi}{1+r_\Lambda} \;. 
\end{eqnarray}

For $r_\Lambda = 1$, one finds $A \left( {\tilde \phi} \right) = 2 \cos \left( \frac{\tilde \phi}{2} \right)$ and $B \left( {\tilde \phi} \right) = 0$. Therefore, the condition $\frac{\partial V_2}{\partial {\tilde \psi}} =0$ leads to the two curves ${\tilde \psi}_s =0$ (a straight line connecting the minimum ${\cal O}$ to the maximum ${\cal M}$), and ${\tilde \phi} = \pi$  (a straight line connecting the two saddle points). Along the first curve, the potential reads $V_1 = 1 - \cos \left( \frac{\tilde \phi}{2} \right)$, and so the trajectory is stable in its first half close to the minimum). Along the second curve, the potential is constant, $V_1 = 1$, and $\frac{\partial^2 V_2}{\partial {\tilde \psi}^2} =0$. This can be understood as the limit  $r_f \rightarrow  1$ of the case $r_\Lambda = \frac{1}{r_f^2}$ shown in the bottom-left panel of Figure \ref{fig:trajectories}.

Let us now discuss the $r_\Lambda \neq 1$ case. We write the solutions of  $\frac{\partial V_2}{\partial {\tilde \psi}} =0$ as the two curves satisfying 
\begin{equation}
\sin \, {\tilde \psi}_s = \pm \frac{B \left( {\tilde \phi} \right)}{\sqrt{A^2 \left( {\tilde \phi} \right)  + B^2 \left( {\tilde \phi} \right)  }} \;\;,\;\;  \cot \, {\tilde \psi}_s = \frac{A \left( {\tilde \phi} \right)}{B \left( {\tilde \phi} \right)} \;. 
\end{equation}
The curve with the $+$ sign connects the minimum ${\cal O}$ with the saddle point ${\cal S}_A$ for $r_\Lambda <1$, and with the saddle point  ${\cal S}_B$ for $r_\Lambda >1$.~\footnote{We recall that the model is symmetric under $r_f \rightarrow \frac{1}{r_f}$ and $r_\Lambda \rightarrow \frac{1}{r_\Lambda}$. In the current case, $r_f = 1$, the symmetry identifies  $r_\Lambda$ with  $\frac{1}{r_\Lambda}$. 
We see that the saddle points interchange role in this identification.}  In both cases the variable ${\tilde \phi}$ ranges from $0$ to $\pi$, and the effective potential along the trajectory evaluates to  
\begin{equation}
V = 1 - \frac{\sqrt{1+ r_\Lambda^2 + 2 \, r_\Lambda \cos {\tilde \phi}}}{1+r_\Lambda} \;. 
\end{equation} 
Since the potential is always $<1$, the curve is a stable valley (see eq. (\ref{rf1-AB})). 

The curve with the  $-$ sign instead connects the maximum ${\cal M}$ to the saddle point  ${\cal S}_B$ for $r_\Lambda <1$, and with the saddle point  ${\cal S}_A$ for $r_\Lambda >1$. This curve in an unstable crest. The behavior and the stability of the two curves agree with those discussed in the main text for $r_f > 1$  (cf.  the top-left panel  of  Figure \ref{fig:trajectories} for $r_\Lambda < 1$, and the bottom-right panel of  Figure \ref{fig:trajectories} for $r_\Lambda > 1$), showing that the $r_f \rightarrow 1$ limit is continuous. 

\subsection{Very small or large $r_\Lambda$} 

These are the regimes in which one term in the potential (\ref{V-start}) is strongly dominant over the other term. The condition $\frac{\partial V}{\partial \psi} =0$ sets this term to a negligible value. Then one is left with the other term, evaluated along the trajectory set by the first term. This reproduces the potential of Natural Inflation. 

To be specific, the  very small (respectively, large) $r_\Lambda$ regime requires $r_\Lambda \ll \frac{1}{r_f^2} ,\,  \frac{1}{r_f^4}$ (respectively,  $r_\Lambda \gg \frac{1}{r_f^2} ,\,  \frac{1}{r_f^4}$). We start the discussion with the very small $r_\Lambda$ regime. 
In this limit, the left hand side of eq.  (\ref{minimumPsi}) can vanish only if the the argument of the first sine is at most of ${\rm O } \left( r_f^2 r_\Lambda \right) \ll1$. This implies
\begin{equation}
{\tilde \psi}_s = \frac{r_f^4 \, r_\Lambda}{1+ r_f^4 r_\Lambda} \, {\tilde \phi} + {\rm O } \left( r_f^3 \, r_\Lambda \right) \,. 
\end{equation} 
and, as a consequence, the argument of the second sine in  eq.  (\ref{minimumPsi}) is $r_f \, {\tilde \phi} + {\rm O } \left( r_f^4 r_\Lambda \right)$. Using this information in the potential (\ref{V2}), we learn that 
\begin{equation}
V = \frac{\Lambda^4}{1+r_\Lambda} \left[ 1 - \cos \left( r_f \, {\tilde \phi} \right) \right] + {\rm O } \left( r_f^4 r_\Lambda^2 \right) \;\;. 
\end{equation} 
Namely, up to negligible corrections, 
\begin{equation}
V \simeq \Lambda_2^4 \left[ 1 - \cos \left( r_f \, {\tilde \phi} \right) \right] \;\;,\;\; r_\Lambda \ll \frac{1}{r_f^2} \,,\; \frac{1}{r_f^4} \,.  
\end{equation} 

A completely analogous computation shows that 
\begin{equation}
V \simeq \Lambda_1^4 \left[ 1 - \cos \left( \frac{{\tilde \phi}}{r_f} \right) \right] \;\;,\;\; r_\Lambda \gg \frac{1}{r_f^2} \,,\; \frac{1}{r_f^4} \,.  
\end{equation}

\section{Cosmological perturbations in the $2$ field model}
\label{sec:perts}

We start by computing the cosmological scalar perturbations for a generic model of two canonically normalized real scalar fields, characterized by the action
\begin{equation}
S = \int d^4 x \sqrt{-g} \left\{ \frac{M_p^2}{2} R - \frac{\left( \partial \varphi_1 \right)^2  +  \left( \partial \varphi_2 \right)^2}{2} - V \left( \varphi_1, \varphi_2 \right)  \right\} \,. 
\end{equation} 
We work in conformal time $\tau$ and in spatially flat gauge: 
\begin{equation}
d s^2 = a^2 \left( \tau \right) \left[ - \left( 1 + 2 \phi \right) d \tau^2 + 2 \partial_i B d x^i d \tau + \delta_{ij} d x^i dx^j \right] 
\;. 
\label{line}
\end{equation}

Moreover, we decompose the scalar fields as a $\tau-$dependent background component plus perturbations
\begin{equation}
\varphi_i = {\phi}_i \left( \tau \right) + \delta \phi_i (\tau,\vec x)
=  {\phi}_i \left( \tau \right) + \int \frac{d^3 k}{\left( 2 \pi \right)^{3/2}} \, {\rm e}^{i  \vec{k} \cdot \vec{x}} \, \frac{Q_i \left( \vec{k} \right)}{a \left( \tau \right)} \,. 
\label{fields}
\end{equation}

We insert (\ref{line}) and (\ref{fields}) into the starting action, and obtain the quadratic order action in the perturbations $S_2$.  We then integrate out the non-dynamical fields $B$ and $\phi$.~\footnote{The fields $\phi \propto \delta g_{00}$ and  $B \propto \delta g_{0i}$ enter in $S_2$ without time derivatives, and so they do not add any extra dynamical degree of freedom. It is customary (see for instance \cite{Maldacena:2002vr}) to integrate them out (i) by evaluating the constraint equations $\frac{\delta S_2}{\delta \phi} = \frac{\delta S_2}{\delta B} =0$, (ii) by solving them expressing the nondynamical variables $\phi$ and $B$ in terms of the dynamical variables $Q_1$ and $Q_2$, and (iii) by  inserting these expressions back into the action $S_2$. In this way $S_2$ becomes an action in terms of the dynamical fields only.} Doing so, one obtains 
\begin{eqnarray}
&& S_2 =  \int d \tau d^3 k \, {\cal L}  \;\;,\;\; 
{\cal L } = \frac{1}{2} \left[ \vert Q_i' \vert^2 - k^2  \vert Q_i \vert^2 - Q_i^* \, M_{ij}^2 Q_j \right] \,, \nonumber\\ 
&& M_{ij}^2 \equiv \frac{{\phi}_i' { \phi}_j'}{M_p^2} \left[ 3 - \frac{a^2}{a'^2} \, \frac{{ \phi}_k' {\phi}_k'}{2 M_p^2} \right] + \frac{a^3}{M_p^2 a'} \left( { \phi}_i' V_{,j} +  V_{,i}  { \phi}_j' \right)  \nonumber\\
&& \quad\quad\quad + a^2 V_{,ij} + \delta_{ij} \left[ \frac{{ \phi}_k' { \phi}_k'}{2 M_p^2} - \frac{2 a'^2}{a^2} \right] \,, 
\label{S2}
\end{eqnarray} 
where $V_{,i} \equiv \frac{\partial V}{\partial \varphi_i}$, and prime denotes differentiation with respect to $\tau$. We note that the variables $Q_i$ are the canonical variables of the problem, and that their conjugate momenta are 
\begin{equation}
\Pi_i \equiv \frac{\partial {\cal L}}{\partial Q_i'} = Q_i^{\dagger '} \;. 
\end{equation} 

We further decompose the Fourier modes $Q_i$ into mode functions times annihilation/creation operators:
\begin{equation}
Q_i \left( \vec{k} \right) \equiv {\cal D}_{ij} \left( k \right) a_j \left( \vec{k} \right) +  {\cal D}_{ij}^* \left( k \right) a_j^\dagger \left( - \vec{k} \right) \;. 
\end{equation} 
The mode functions satisfy the classical equations of motion 
\begin{equation}
{\cal D}'' + \Omega^2 \, {\cal D} = 0 \;\;,\;\; \Omega^2 \equiv k^2 \, {\bf 1} + M^2 \;, 
\label{eom}
\end{equation}
where $\bf{1}$ is the identity operator, and $M^2$ the squared mass matrix defined in (\ref{S2}). To  quantize the system, we impose standard Equal Time Commutation Relations among the fields and their conjugate variables, and the standard algebra on the annihilation/creation operators: 

\begin{eqnarray}
&& \left[ Q_i \left( t, \vec{x} \right) , \Pi_j \left( t, \vec{y} \right) \right] = i \delta_{ij} \, \delta^{(3)} \left( \vec{x} - \vec{y} \right) 
\;, \nonumber\\ 
&& \left[ a_i \left( \vec{k} \right), a_j^\dagger \left( \vec{p} \right) \right] = \delta_{ij} \, \delta^{(3)} \left( \vec{k} - \vec{p} \right) 
\;. 
\label{quantization}
\end{eqnarray} 

To ensure these relations,  we impose 
\begin{eqnarray}
&& {\cal D} \, {\cal D}^{\dagger'} - {\cal D}^* {\cal D}^T = - i  \;, \nonumber\\ 
&& {\cal D} \, {\cal D}^\dagger = {\rm real} \;\;,\;\; 
{\cal D}' \, {\cal D}^{\dagger '} = {\rm real} \,, 
\label{initialcond}
\end{eqnarray} 
at some given initial time $\tau_0$. The first condition in (\ref{initialcond}) ensures that both relations (\ref{quantization}) can be imposed at $\tau_0$. Enforcing all three of (\ref{initialcond}) at $\tau_0$ ensures that these conditions continue to hold at  later times (as can be immediately verified taking their time derivatives and using (\ref{eom})). This ensures that the first relation in  (\ref{quantization}) holds at all times. 

We impose the initial conditions when a mode is deeply inside the horizon. Specifically, the physical momentum $\frac{k}{a \left( \tau_0 \right)}$ should dominate the dispersion relation of the mode. The entries in the mass matrix (\ref{S2}) are of ${\rm O } \left( a^2 H^2 ,\, a^2 m_i^2 \right)$, where $m_{1,2}$ are the two mass eigenvalues of $M$.  Therefore, we ensure that $\frac{k}{a \left( \tau_0 \right) } \gg H ,\, m_i$. Neglecting slow roll corrections, $a = - \frac{1}{H \tau}$ during inflation, and so we impose
\begin{equation}
\vert \tau_0 \vert  \! \gg  \! \frac{1}{k} , \frac{m_i}{H \, k} \;\Rightarrow \; \Omega^2 \left( \tau_0 \right) = 
 k^2  \left[ {\bf 1} + {\rm O }  \! \left( \frac{1}{k^2 \, \tau_0^2} ,\, \frac{m_i^2}{k^2 H^2 \tau_0^2} \right) \! \right] \,. 
\end{equation} 

Using this approximated expression for $\Omega$, we see that the equations of motion (\ref{eom}) admit the early time solution ${\cal D}_{ij,{\rm early}} \simeq \frac{{\rm e}^{-i k \tau}}{\sqrt{2 k}} \, \delta_{ij}$. So, we impose, 
\begin{equation}
{\cal D}_{ij} \left( \tau_0 \right) = \frac{1}{\sqrt{2 k}} \, \delta_{ij} \;,\;\; {\cal D}_{ij}' \left( \tau_0 \right) = - i \sqrt{\frac{k}{2}} \, \delta_{ij} \,, 
\label{in-cond}
\end{equation}
which satisfy all three of  (\ref{initialcond}). Namely, in the deep sub-horizon regime the mass matrix $M$ is irrelevant, and the two eigenmodes of the system behave as decoupled massless modes in Minkowski space-time. The solution just written is the standard adiabatic vacuum solution for these decoupled  modes. 

We are interested in observables that are linear in the two scalar fields, ${\cal O} \left( \tau , \vec{x} \right) =  c_{i} \left( t \right) \, \delta \phi_i \left( t ,\, \vec{x} \right)$, where the (real) coefficients $c_i$ can depend on background quantities (and, hence, on time). We assume that we have a set of such observables $\left\{ {\cal O}^{(1)} ,\,  {\cal O}^{(2)} ,\, \dots ,\, 
  {\cal O}^{(n)} \right\}$, so that 
\begin{equation}
{\cal O}^{\left(\alpha\right)} \left( \tau , \vec{x} \right) =  c_{\alpha i} \left( t \right) \, \delta \phi_i \left( t ,\, \vec{x} \right) \,, 
\label{observables}
\end{equation}
and we compute the two point correlation function among pairs of them 
\begin{eqnarray}
&& {\cal C}_{\alpha \beta} \left( t ,\, \vec{x} ,\, \vec{y} \right) \equiv 
\left\langle {\cal O}^{(\alpha)} \left( t ,\, \vec{x} \right)  {\cal O}^{(\beta)} \left( t ,\, \vec{y} \right)  \right\rangle \nonumber\\ 
&& \quad\quad = \int_0^\infty \frac{d k}{k} \; \frac{\sin \left( k r \right)}{k r} \, P_{\alpha \beta} \left( t, k \right) \;\;,\;\; 
r = \vert \vec{x} - \vec{y} \vert \,, \nonumber\\ 
\end{eqnarray} 
where  we have used the fact that (thanks to statistical isotropy) the mode functions ${\cal D}$ do not depend on the direction of $\vec{k}$, and where we have introduced the power spectra 
\begin{equation}
P_{\alpha \beta}  \left( \tau, k \right) \equiv 
\frac{k^3}{2 \pi^2 a^2 \left( \tau \right)} 
\left[ c \left( \tau \right) \, {\cal D} \left( \tau , k \right)  {\cal D}^\dagger \left( \tau , k \right) c^T \left( \tau \right) \right]_{\alpha \beta} \;, 
\label{PS} 
\end{equation}
(we note that this expression is symmetric and real). 

In shorts, for any set of observables (\ref{observables}), we compute the self- and cross-power spectra by imposing the initial conditions (\ref{in-cond}), by evolving (\ref{eom}), and by evaluating (\ref{PS}). 

In this work, we have two scalar fields $\varphi_1 = \theta$ and $\varphi_2 = \rho$, and we are interested in the power in the adiabatic ${\cal R}$ and entropy ${\cal S}$ modes, that, in spatially flat gauge, read (see \cite{Malik:2008im} for a review) 
\begin{equation}
\left( \begin{array}{c} {\cal R} \\ {\cal S} \end{array} \right) = 
\frac{\cal H}{\vert \vec{\phi}' \vert} \left( \begin{array}{cc} \cos \theta & \sin \theta \\ - \sin \theta & \cos \theta \end{array} \right) 
\, \left( \begin{array}{c} \delta \phi_1 \\ \delta \phi_2 \end{array} \right) ,  
\label{RS-rotation}
\end{equation} 
where we have defined 
\begin{equation}
\cos \theta \equiv \frac{\phi_1'}{\vert \vec{\phi}' \vert} \;,\; 
\sin \theta \equiv \frac{\phi_2'}{\vert \vec{\phi}' \vert} \;,\; 
\vert \vec{\phi}' \vert \equiv \sqrt{ \phi_1^{' 2} +  \phi_2^{' 2}  } \;. 
\end{equation}

\begin{figure}
\centerline{
\includegraphics[width=0.45\textwidth,angle=0]{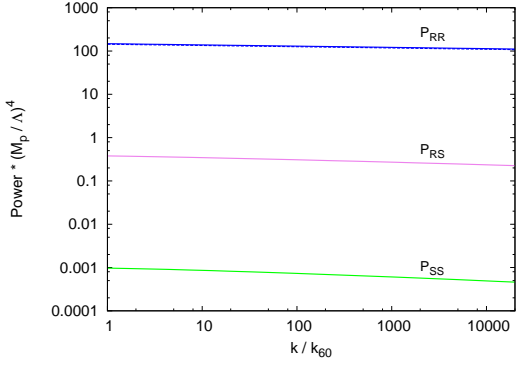}
}
\caption{Solid lines: Auto- and cross- power spectra for the adiabatic and entropy scalar perturbations in Aligned Natural Inflation, obtained from a numerical evolution with the model (\ref{V-start}), for the parameters $r_\Lambda = 0.3 ,\, r_f = 1.5 ,\; r_g = 1 ,\; f_\phi = 5 \sqrt{2} M_p ,\; f_\psi = \frac{M_p}{10 \sqrt{2} } \,$. $k_{60}$ is the comoving momentum of the mode that exits the horizon $60$ e-folds before the end of inflation. Inflation starts next to the saddle point ${\cal S}_B$, and ends when the trajectory becomes unstable in the heavy direction. Dashed line: Slow roll result (\ref{slowroll}) for $P_{\cal RR}$ in the single field approximation (assuming ${\cal S} = 0$). 
}
\label{fig:PS}
\end{figure}

The solid lines of  Figure (\ref{fig:PS}) show the auto- and cross- spectra of the adiabatic and entropy modes for an illustrative choice of parameters, with inflation taking place along  the high-altitude plateau connected to the saddle point ${\cal S}_B$. The adiabatic mode corresponds to perturbations along the light inflationary direction, while the entropy mode corresponds to the much heavier orthogonal direction (the alignment parameter is $\alpha = \frac{1}{100}$). As expected, the entropy mode is significantly smaller than the adiabatic one, and we can simply disregard it, with very good accuracy. Starting from the quadratic action (\ref{S2}), performing the rotation (\ref{RS-rotation}), and setting ${\cal S} = 0$, we obtain the theory for the single ${\cal R}$ perturbation, in the limit in which ${\cal S}$ can be disregarded. As expected, we explicitly verified that, in this limit, the quadratic action for  ${\cal R}$ reproduces that of a single field with a potential equal to the potential of the $2-$field  model along the inflationary trajectory. Specifically, setting ${\cal S}=0$, we obtain 
\begin{eqnarray}
{\cal R} &\equiv& \frac{v}{z} \;\;,\;\; z \equiv \frac{a \, \vert \vec{\phi}' \vert }{H} \nonumber\\ 
S_2 &=& \frac{1}{2} \int d \tau d^3 k \left[ \vert v' \vert^2 - \left( k^2 - \frac{z''}{z} \right) \vert v \vert^2 \right] \,. 
\end{eqnarray} 
which is the standard action \cite{Mukhanov:1990me} for the Mukhanov-Sasaki variable \cite{Mukhanov:1985rz,Sasaki:1986hm} in the single field case. 

One may then use the standard slow roll relations for the perturbations, 
\begin{equation}
P_{\cal R} = \frac{H^2}{8 \pi^2 \epsilon M_p^2} \;\;,\;\; n_s - 1 = 2 \eta - 6 \epsilon \;\;,\;\; r = 16 \, \epsilon \;, 
\label{slowroll}
\end{equation}
with the slow roll parameters evaluated along the inflationary trajectory 
\begin{eqnarray}
\epsilon &=& \frac{M_p^2}{2} \: \frac{\vert \vec{\nabla} V \vert^2}{V^2} 
\approx \frac{3 \, \vert \vec{\dot{\phi}} \vert^2}{2 V} \;, \nonumber\\ 
\eta &=& M_p^2 \, \frac{ {\hat \nabla}_i V \: V_{,ij} \:  {\hat \nabla}_j V }{ V } \approx 
M_p^2 \, \frac{V_{,ij}}{V} \, \frac{\dot{\phi}_i}{\vert \vec{\dot{\phi}} \vert } \, 
 \frac{\dot{\phi}_j}{\vert \vec{\dot{\phi}} \vert } \,.  
\label{eps-eta}
\end{eqnarray} 

The dotted line in Figure \ref{fig:PS} shows the power spectrum for ${\cal R}$ obtained through the slow-roll relation 
(\ref{slowroll}), with the slow roll parameter $\epsilon$ evaluated through (\ref{eps-eta}). This differs by about $2\%$ from the exact result indicated by the solid line. The discrepancy is consistent with the accuracy of the slow-roll approximation (in this run, $\eta \simeq 1.5\%$).

\section{An incorrect simplification ($\psi=0$)} 
\label{app:incorrect}

In this Appendix we warn against a  simplification that in general provides incorrect phenomenological results. The transformation (\ref{rotation}) rotates (up to negligible ${\rm O } \left( \alpha^2 \right)$ corrections) the original axions $\theta$ and $\phi$ of the starting potential (\ref{V-start}) into the linear combinations $\phi$ and $\psi$. Such fields are, respectively, the light and heavy directions of the inflaton potential close to the minimum at $\theta = \rho =0$. For this reason, the inflationary evolution next to the origin occurs along the $\phi$ direction. One therefore may be tempted to 
simply set $\psi=0$ all throughout inflation. This would lead to the single field potential (\ref{V2}) with ${\tilde \psi} =0$.

\begin{figure}
\centerline{
\includegraphics[width=0.4\textwidth,angle=0]{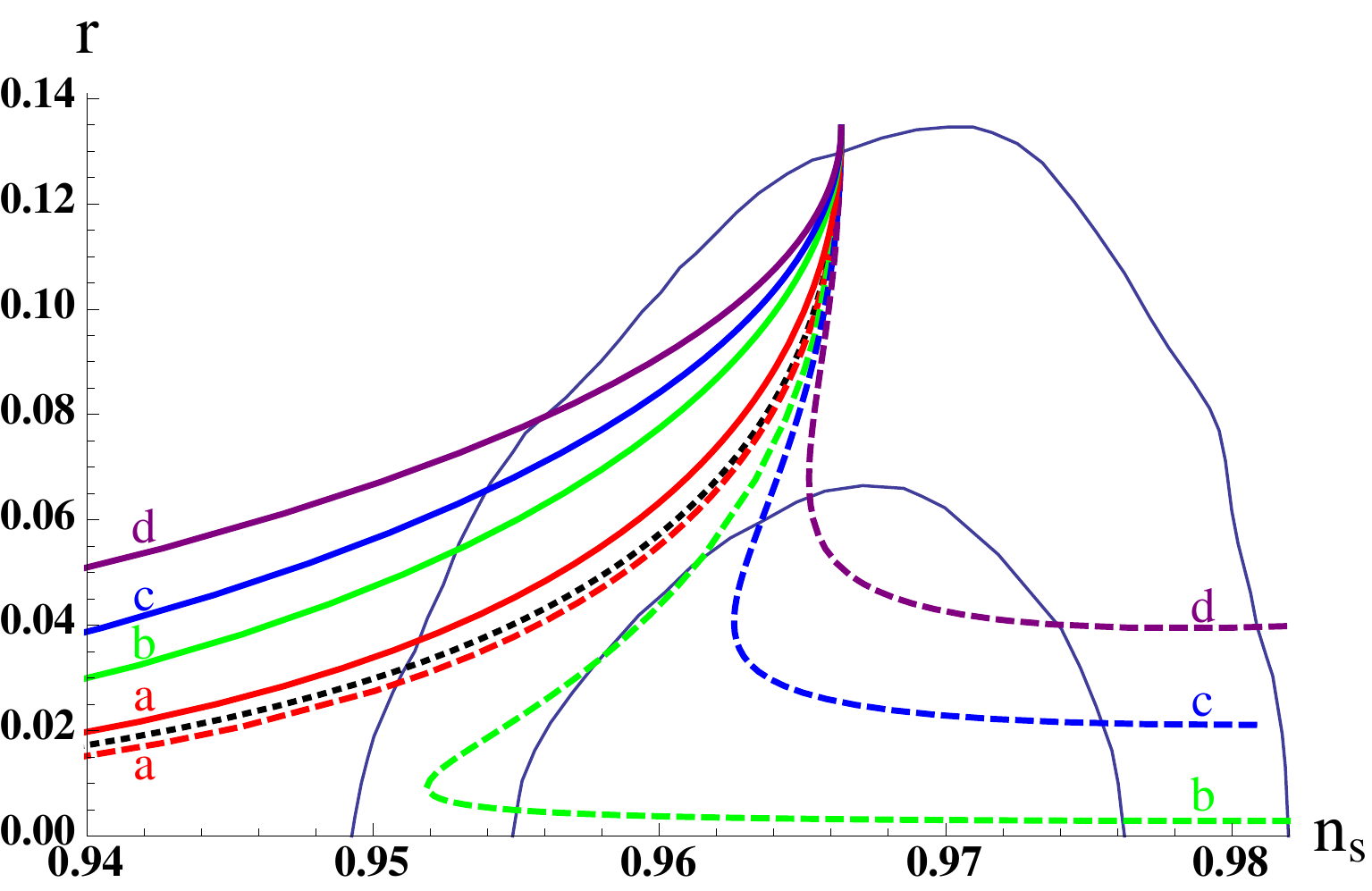}
}
\caption{Solid lines: correct phenomenological results; dashed lines: incorrect results obtained from incorrectly setting 
$\psi=0$. The labels on the curves correspond to the following values of $\left\{ r_f ,\, r_\Lambda \right\}$: a $= \left\{ 1.5 ,\, 3 \right\} \,$; b $= \left\{ 1.5 ,\, 0.05 \right\} \,$; 
 c $= \left\{ 1.65 ,\, 0.05 \right\} \,$; 
 d $= \left\{ 1.8 ,\, 0.05 \right\} \,$. For reference, the prediction of Natural Inflation is shown as the black dotted line. 
 }
\label{fig:psi=0}
\end{figure}

In Figure \ref{fig:psi=0} we show with dashed lines the phenomenological results obtained under this incorrect assumption, for four different choices of parameters. As a comparison, we also show with solid lines the correct phenomenological results, obtained from an exact numerical evolution. We see that the approximated results shown are in reasonable agreement with the correct ones only in the top part of the lines, where $r$ is greatest. 

Each line is obtained by varying only $f_\phi$ and keeping the other parameters constant. The common point of all the lines corresponds to $f_\phi \rightarrow \infty$. In this limit, ${\tilde \phi} = \frac{\phi}{f_\phi}$ only probes the immediate vicinity of the minimum, where the model becomes identical to massive chaotic inflation (in practice, only the first quadratic term of the Taylor expansion of the potential around the minimum matters). As  setting $\psi = 0$  is a good approximation close to the minimum, the approximate dashed lines approach the correct solid lines in that limit. However, as $f_\phi$ decreases, the system probes the potential far from minima, where the $\psi=0$ assumption is incorrect. This explains why setting $\psi=0$ does generally not provide the correct phenomenological results for the model.

\end{document}